\documentclass[twocolumn]{amsart}%

\usepackage{graphicx,natbib,pdfpages}
\usepackage[margin=2.4cm]{geometry}
\usepackage{color}
\newcommand{\tc}{\text{:}} % tc=tight colon, for use in math mode in Newick trees
\begin{document}

\title[Testing Coalescent Simulators]{Testing Multispecies Coalescent Simulators using Summary Statistics}
\author{Elizabeth S. Allman}
\address{Department of Mathematics and Statistics, University of Alaska Fairbanks, Box 756660, Fairbanks, AK, 99775-6660}
\email{e.allman@alaska.edu}

\author{Hector Ba\~nos}
\address{Department of Mathematics and Statistics, University of Alaska Fairbanks, Box 756660, Fairbanks, AK, 99775-6660}
\curraddr{School of Mathematics, Georgia Institute of Technology, 686 Cherry Street, Atlanta, GA 30332-0160}
\email{hbassnos@gmail.com}

\author{John A. Rhodes}
\address{Department of Mathematics and Statistics, University of Alaska Fairbanks, Box 756660, Fairbanks, AK, 99775-6660}
\email{j.rhodes@alaska.edu}

\begin{abstract}As genomic scale datasets motivate research on species tree
  inference, simulators of the multispecies coalescent (MSC) process
  are essential for the testing and evaluation of new inference
  methods. However, the simulators themselves must be tested to ensure
  they give valid samples from the coalescent process. In this work we
  develop several statistical tools using summary statistics to evaluate 
  the fit of a simulated gene tree sample to the MSC model. Using these
  tests on samples from four published simulators, we uncover flaws in
  several. The tests are implemented as an R package, so that both
  developers and users will be able to easily check proper performance
  of future simulators.
 \end{abstract}

\maketitle

\section{Introduction}

With the increasing availability of genomic-scale datasets, comprised
of sequences from many genetic loci, it has become clear that
individual gene trees inferred for the same taxa are often
discordant. While gene tree inference error may be a factor, there are
also many biological processes that could lead to this discordance. Of
these, \emph{incomplete lineage sorting} as described by the
\emph{multispecies coalescent} (MSC) model is often viewed as the most
fundamental. The MSC may thus be thought of as the null model of
discordance, to be considered before further complications such as
hybridization, lateral gene transfer, gene duplication and loss,
and/or population structure, are invoked \citep{Degnan2018}. While
inference of species trees under the MSC is still challenging, it can
now be performed in a variety of ways
\citep{Liu2008,Heled2010,Chifman2014,Vachaspati2015,Zhang2018}.

Evaluating the performance of any approach to species tree inference
under the MSC, though, requires testing on simulated datasets, so that
the true species tree is known and accuracy can be judged. Thus MSC
software for simulating the formation of gene trees within a species
tree or network, often paired with software for simulating the
evolution of sequences along the gene trees, plays a critical role in
advancing methodology. Unfortunately, as we demonstrate below, some of
the available simulators do not produce valid MSC samples, although
this may not be apparent even to knowledgeable users.

Testing a coalescent simulator for correctness is not simple, as the
theoretical distribution of gene trees gives positive probability to
all gene tree topologies, with the probability density for metric gene
trees having a quite complicated dependence on the species tree branch
lengths.  ``Straightforward'' comparison of a sample to the
theoretical distribution is simply not a practical approach. This has
led some developers to validate their software by comparing output to
other simulators, rather than to theoretical predictions
\citep{simphy}.  Here we introduce several testing tools, based on
summary statistics that capture either topological or metric
information. Implemented in an R package, \texttt{MSCsimtester}, these
can be applied to gene trees samples from any simulator to study
whether its output is in accord with the MSC.  Although examining such
summary statistics cannot give an ironclad guarantee of correctness,
we believe they are likely to uncover most problems.

\ 

When we applied the \texttt{MSCsimtester} tools to output from four
well-known MSC simulators, we discovered that only one of the four,
SimPhy \citep{simphy}, behaved as we expected. The documentation for
one, Phybase \citep{phybase}, was not sufficiently explicit on
specifying its input and our initial interpretation was incorrect. A
third, Hybrid-lambda \citep{Hybrid-lambda} passes our tests for gene
tree topologies, but samples metric gene trees incorrectly. The
fourth, Mesquite \citep{Mesquite}, produced samples with neither gene
tree topologies nor metric properties in accord with the
model. However, we found website documentation of only the
Hybrid-lambda issue (which we had uncovered in a preliminary version
of this work).

While we notified the authors of these simulators of the problems in
advance of this publication, the larger community should be aware of
the need to interpret results of previous simulation work with them
cautiously. We also suggest that users of other simulators, and
developers of new ones, test them with the \texttt{MSCsimtester}
tools, which are available at the
\texttt{jarhodesuaf.github.io/software.html} website.

\section{New Approaches}

We give an informal description of the MSC to introduce the summary statistics we focus upon.

\begin{figure}
	\includegraphics[scale=.21]{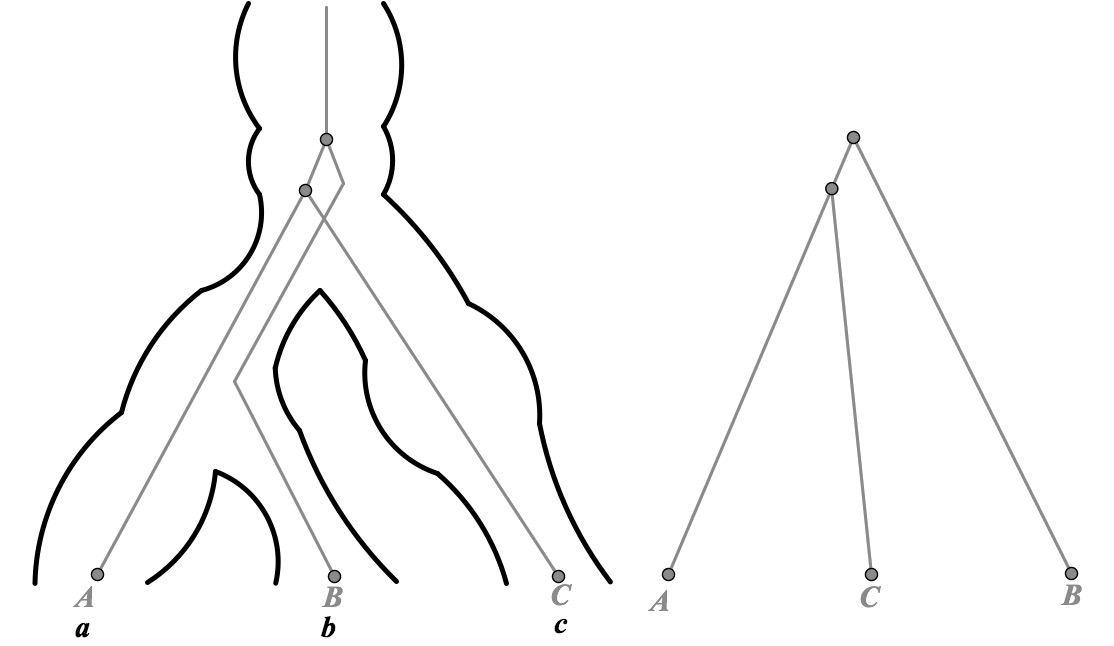}\hspace{.5cm}  
	\caption{(Left) A metric species tree
          $((a\tc\ell_a,b\tc\ell_b)\tc\ell_v,c\tc\ell_c)$, with
          population sizes depicted by widths of edges. With one
          lineage $A,B,C$ sampled from each species, a metric gene
          tree depicting ancestral lineages forms within it. (Right)
          The same metric gene tree depicted in more standard
          fashion.} \label{fig::ILS}
\end{figure}

Suppose first that 3 taxa are related by the species tree with
topology $((a,b),c)$. Acknowledging that species are composed of
populations, we depict this by a tree whose edges are `pipes' as in
Figure \ref{fig::ILS}. The length of each pipe is elapsed time
measured in generations, and the width of the pipe represents
population size, which may vary over time and edge. When individual
genes are sampled from the leaves of the species tree, they trace
backwards in time within the species tree until they coalesce at a
common ancestral individual. Coalescence is a random process, which can be thought
of as individual genes on the lineage choosing their `parent'
uniformly at random from those existing in the previous generation, a
panmictic viewpoint. Thus the only population detail of importance
under the MSC is size. Importantly, there is a greater chance of
coalescence when populations are small, but no requirement that
lineages coalesce within any specified finite time.

Often, simplifying assumptions on population sizes are made by
modelers and programmers, such as all populations at all times and on
all branches throughout the tree are a constant $N$. More realistic is
to at least allow different population sizes $N_e$ for each edge
(pipe) $e$ of the species tree. While it would be highly desirable to
be able to simulate gene trees under the MSC using arbitrary
population size functions, current simulators do not makes this easy
on a large tree. Nonetheless, our testing methods accommodate that
generality.

\medskip

When gene trees are produced under the MSC they have two aspects that
can be viewed somewhat separately. One is the metric information,
which is reflected in the distribution of pairwise distances between
two fixed taxa across the gene trees. The second is topology, which is
reflected in the distribution of rooted triple trees on three fixed
taxa displayed on the gene trees. In testing the performance of a
simulator, it is important both metric and topological aspects be
examined.

\subsection{The distribution of pairwise distances on gene trees}
\begin{figure}
        {\includegraphics[width=8cm]{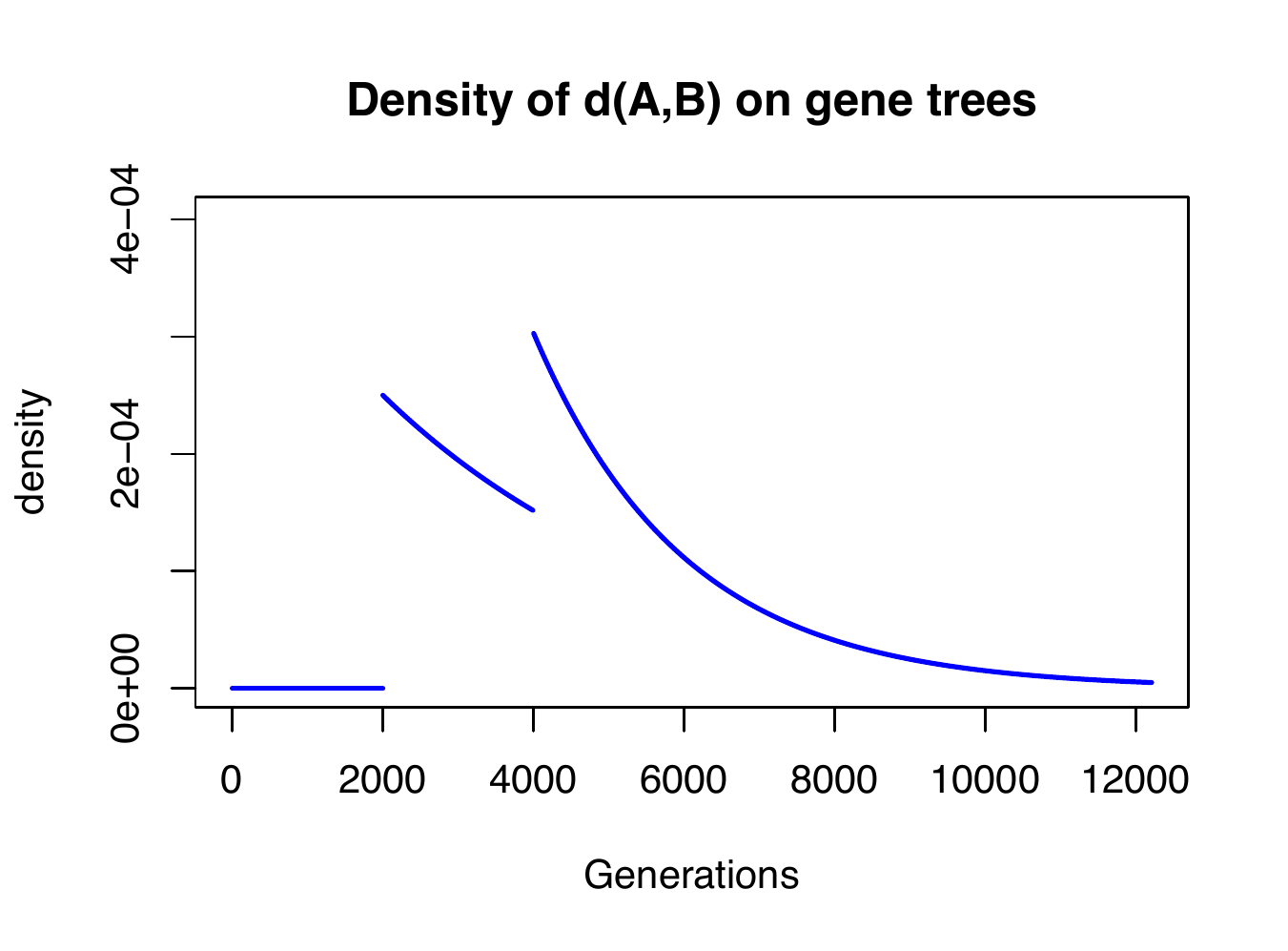}  }
 	 	
	\caption{The plot of the probability density function of $d(a,b)$ on gene trees for  the species tree shown in Figure \ref{fig::ILS}, 
	with all branch lengths equal to 1000 generations.  The population size parameters are $N = 2000$ on the internal edge, 
	and $N = 1000$ for the population ancestral to the root.	
	}\label{fig::distexample}
	
 \end{figure}

Consider now the species tree
$$((a\tc\ell_a,b\tc\ell_b)\tc\ell_v,c\tc\ell_c)$$ of Figure
\ref{fig::ILS}, with root $r$ and $v$ the most recent common ancestor
of $a$ and $b$. Fix constant population sizes $N_{v}$ on the edge
above $v$, and $N_r$ above the root $r$. For the purpose of
illustration, assume $N_r$ is \emph{smaller} than $N_{v}$. Tracking
two genes $A$ and $B$ sampled from $a$ and $b$ backwards, their
lineages cannot coalesce until the reach the population above $v$. On
the branch above $v$ coalescence occurs by a Poisson process at a
constant rate $1/N_{v}$. This leads to the time to coalescence above
$v$ being exponentially distributed, at least for those times more
recent than the root of the tree. At the root of the tree there is a
discontinuous change in the coalescence rate, to the \emph{larger}
value $1/N_{r}$. Coalescence above the root $r$ again occurs with an
exponential distribution, but at a faster rate. Putting together the 3
regions analyzed here (below $v$, between $v$ and $r$, above $r$, we
find the distribution of distances between $a$ and $b$ on gene trees
is a piecewise exponential, such as shown in Figure
\ref{fig::distexample}.

 Note that the discontinuities in the distribution in Figure
 \ref{fig::distexample} occur due to 1) the impossibility of
 coalescence below $v$ and 2) the discontinuity in population size at
 $r$. For larger trees, with more edges leading from the MRCA of a
 pair of taxa to the root, there can of course be more discontinuities
 in the density even when the population size is constant on
 each edge. If population sizes vary on edges, the pieces need not
 come from exponential distributions, but are computable from the
 population size function. Discontinuities in the distribution arise
 whenever there is a discontinuity in the population size. Finally
 none of this depends on the species tree being ultrametric, which of
 course it need not be biologically since edge lengths are in
 generations. The precise form of the pairwise distance distribution,
 and its derivation, is given in the Methods section.

 To test an MSC simulator using this distribution, one can produce a
 large sample of gene trees from a fixed species tree with population
 sizes, and then choosing some pair of taxa compare a histogram of the
 pairwise distances between these taxa across the gene trees to the
 theoretical distribution. This comparison can be done visually, as
 major deviations from the theoretical predictions will be
 obvious. One can further perform a statistical test, such as that
 developed by \citet{Anderson1952}, to compare the empirical
 distribution from the simulator to the theoretical one, giving a
 $p$-value.

\subsection{The distribution of rooted triple topologies on gene trees}

Returning to the species tree of Figure \ref{fig::ILS}, we can ignore
distances on gene trees and instead examine topological
features. Supposing again we have constant population sizes $N_v$ and
$N_r$, the chance that lineages from $a$ and $b$ fail to coalesce in
the edge of length $\ell_v$ between $v$ and $r$ can be computed to be
$e^{-x},$ where $x=\ell_v/N_v$ \citep{Pamilo1988}. Note that a longer
edge (increasing $\ell_v$) and/or a smaller population (decreasing
$N_v$) make coalescence more likely. The quantity $x$ here is the
length of the edge in \emph{coalescent units}, a convenient unit to
address the confounded effects of population size and time. If the
lineages fail to coalesce before the root, then lineages from $a,b,c$
will all be present above $r$, and all three rooted gene topologies
$((a,b),c)$, $((a,c),b)$, and $((b,c),a)$ are equally likely to
form. This leads to the gene tree probabilities
\begin{align*}
\mathbb{P} (((a,b),c))&= 1-2/3e^{-x},\\  
\mathbb{P} (((a,c),b))&= 1/3e^{-x},\\ 
\mathbb{P} (((b,c),a))&=1/3 e^{-x}
\end{align*}
which were derived by \citet{Pamilo1988}. More general formulae,
accommodating changing population sizes, are given in the Methods
section.

Given a large sample of gene trees from an MSC simulator, for any 3
fixed taxa one can tabulate the frequencies of the three rooted
topologies displayed on the gene trees. The simulation can then be
tested in two ways: First, one may find the maximum likelihood
estimator of $x$ from the tabulated frequencies, and compare this to
the theoretical value. Second, one can judge the fit of the empirical
frequencies to the expected ones by calculating, for instance, a
chi-squared statistic and judging it against the $\chi^2$ distribution
with 2 degrees of freedom to determine a $p$-value. The second of
these is sensitive to imbalances between the counts for the two
topologies incongruent with the species tree, while the first is not.

\section{Results and Discussion}

We focused on simulators that 1) allowed easy input of a species tree
either in Newick notation, or graphically, with branch lengths in
generations and 2) allowed a population size to be assigned to each
branch in the species tree independently. The four simulators we
investigated were Mesquite \citep{Mesquite}, Phybase \citep{phybase},
Hybrid-lambda \citep{Hybrid-lambda}, and SimPhy \citep{simphy}. Our
criteria ruled out several other well-known simulators, including ms
\citep{Hudson2002} (which does not provide for easy input of a species
tree in Newick), and PhyloNet \citep{phylonet} (which calls ms and
allows for a Newick tree, but does not allow different population
sizes on different branches).

\begin{figure}[t]
 	\includegraphics[scale=.41]{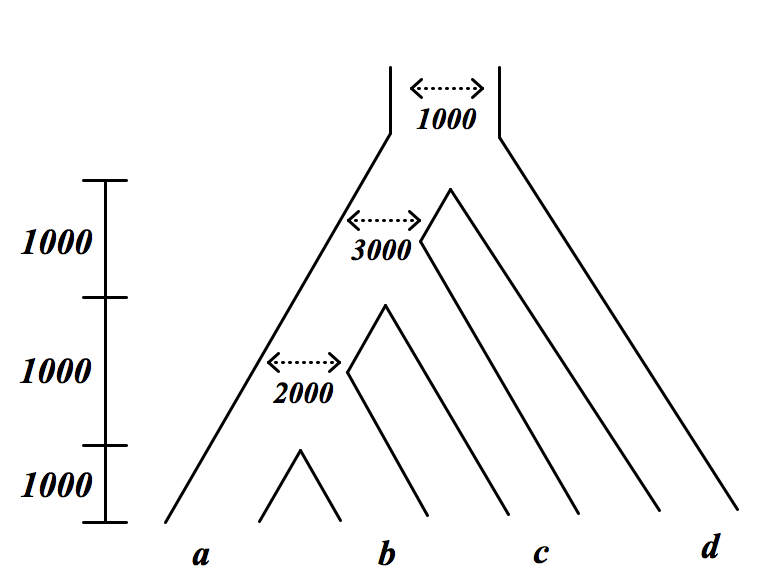} 
	\vskip .1in
 	\caption{4-taxon metric species tree, with constant population sizes on each edge, for which data was simulated.}\label{fig::simtree}
 \end{figure}

\begin{figure*}
\begin{center}
        {   \includegraphics[width=7.5cm]{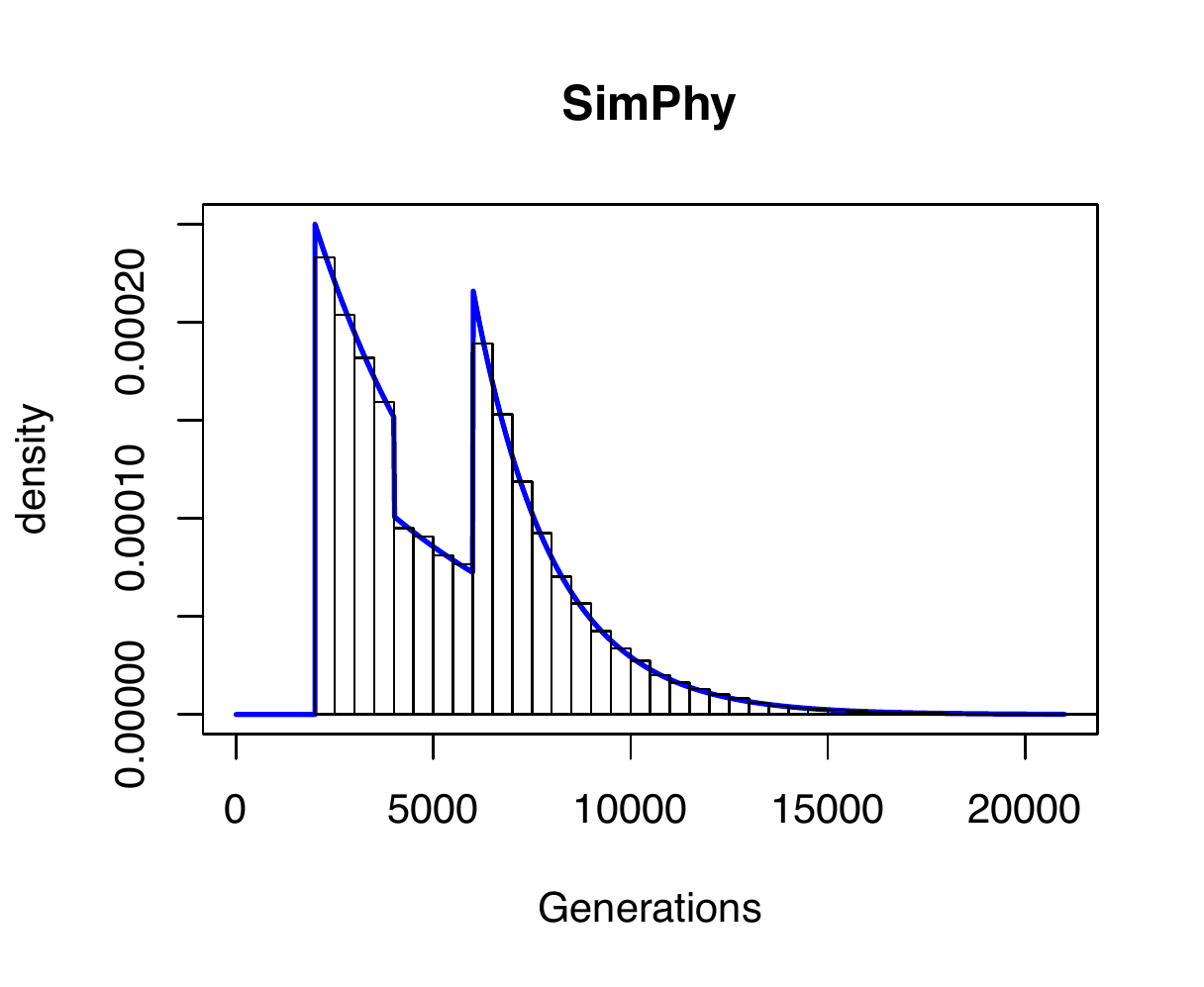}
             \includegraphics[width=7.5cm]{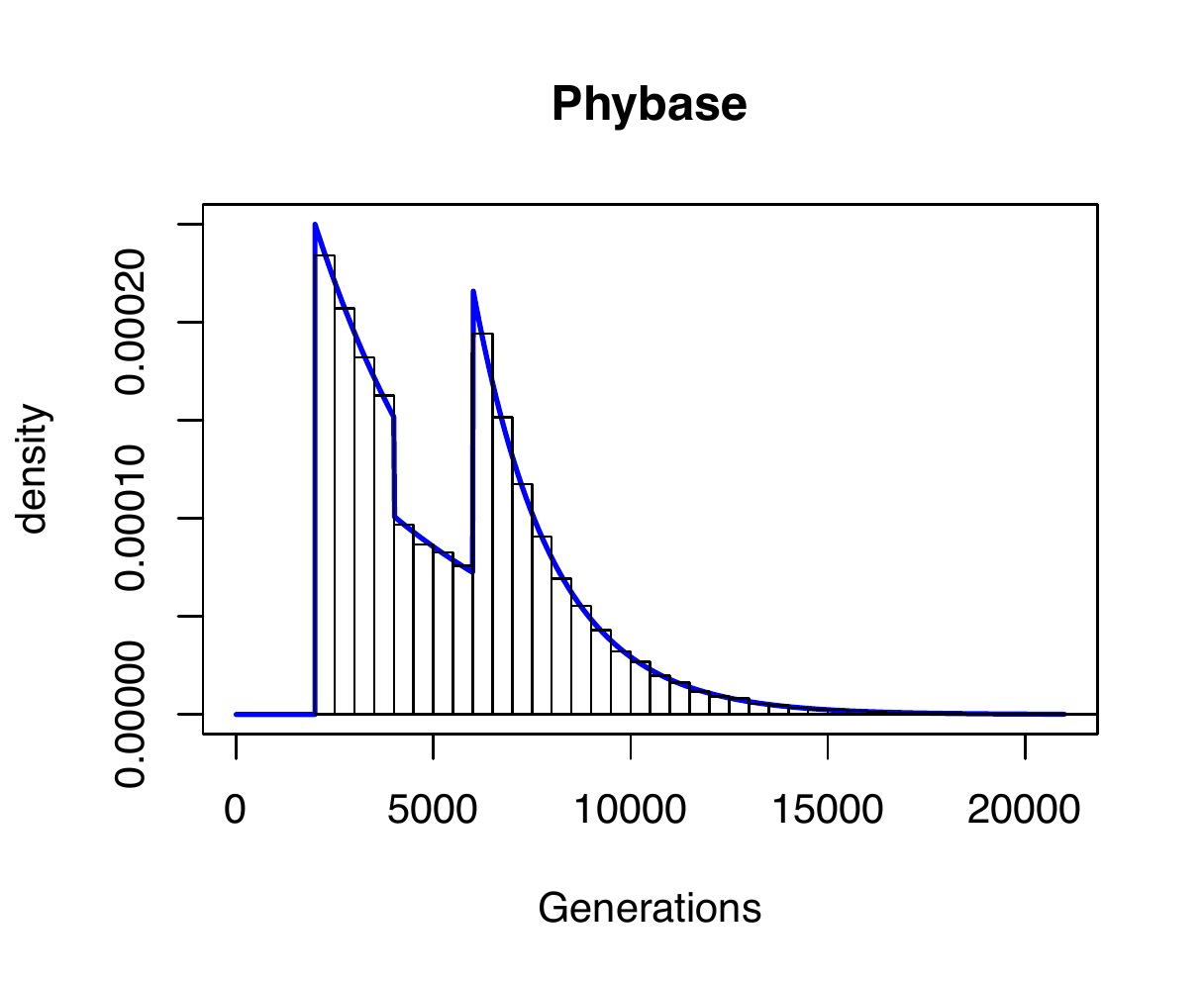}\\
             \includegraphics[width=7.5cm,height=6.15cm]{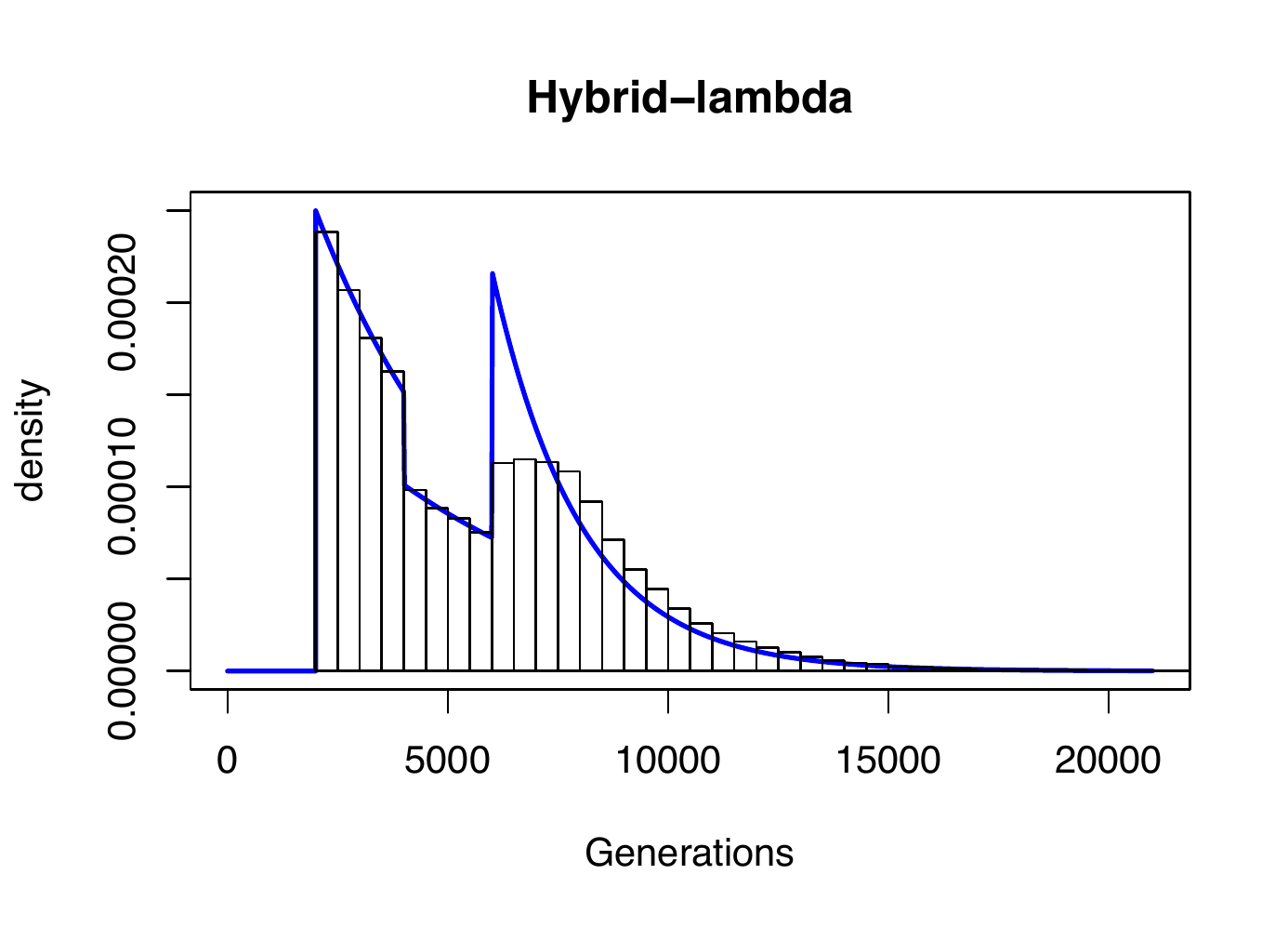}
             \includegraphics[width=7.5cm]{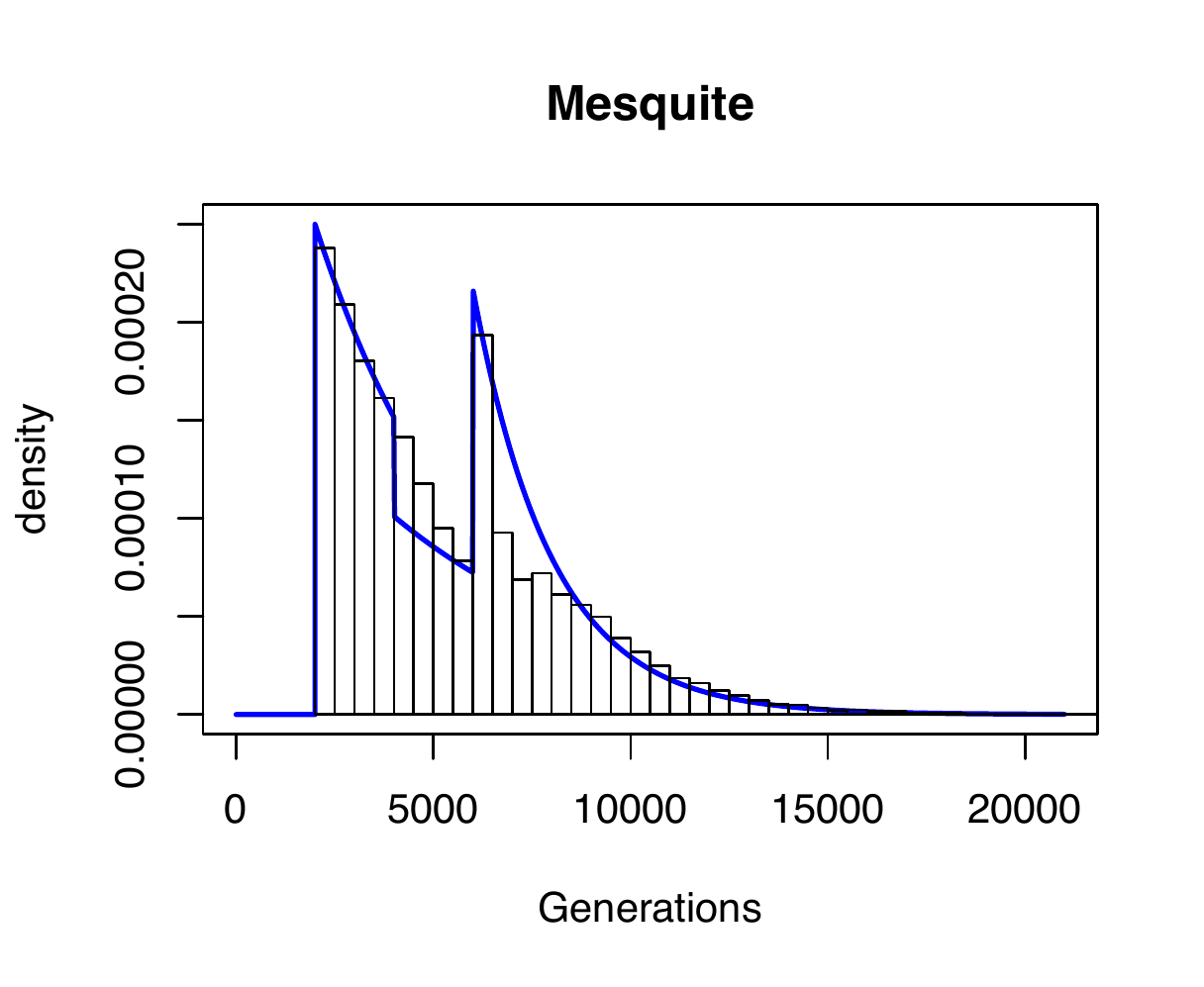}  } 
\end{center}
\caption{Pairwise distance distributions for $d(A,B)$ from 100,000
  simulated gene trees for species tree and population sizes as shown
  in Figure \ref{fig::simtree}. } \label{fig::pairwise}
\end{figure*}

 Mesquite is a popular package with many phylogenetic tools accessible
 through a graphical interface. Phybase is an R package which also
 includes phylogenetic tools beyond coalescent
 simulation. Hybrid-lambda implements the standard coalescent
 model, and the more general $\lambda$-coalescent, on both trees and
 hybridization networks. SimPhy includes a very general model of gene
 tree generation, including gene duplication and loss. Our tests,
 therefore, should not be construed as complete tests of any of this
 software; we test only a standard MSC simulation on a species tree,
 with branch lengths in generations.

We performed tests with a number of species trees, with up to 6 taxa,
using a variety of constant population sizes.  Here we show only
representative examples of this work, using the species tree and
populations depicted in Figure \ref{fig::simtree}. Additional test
results are shown in the supplementary materials, in Figures S2-S21
and Tables S1-S4.

A sample of 100,000 gene trees was simulated with each of the
programs, the values of $d(A,B)$ on them was extracted, and a
histogram produced.  These are shown in Figure \ref{fig::pairwise},
with the theoretical distribution superimposed on them. There is a
good match for SymPhy, as was seen in all our simulations with this
software.  When our initial Phybase simulation did not match
expectations, we learned that species tree branch lengths should be
supplied as $\mu t$, where $t$ is in generations and $\mu$ is a
mutation rate, while population sizes should be specified as
$\theta=4\mu N_d$ where $N_d$ is diploid population size. Taking
$\mu=1$ and $\theta=2N$ with $N$ the haploid population size,
Phybase's sample matched expectations well.  Both the Hybrid-lambda
and Mesquite simulations show pronounced deviations from the
theoretical distribution. However, when the input tree is specified in
coalescent units, Hybrid-lambda does give a sample of gene trees whose
pairwise distances in coalescent units match the theory well; the poor
fit occurs only when the species tree parameters involve separate time
and population values.

\begin{figure*}[t]
\begin{center}
\includegraphics[width=7.5cm]{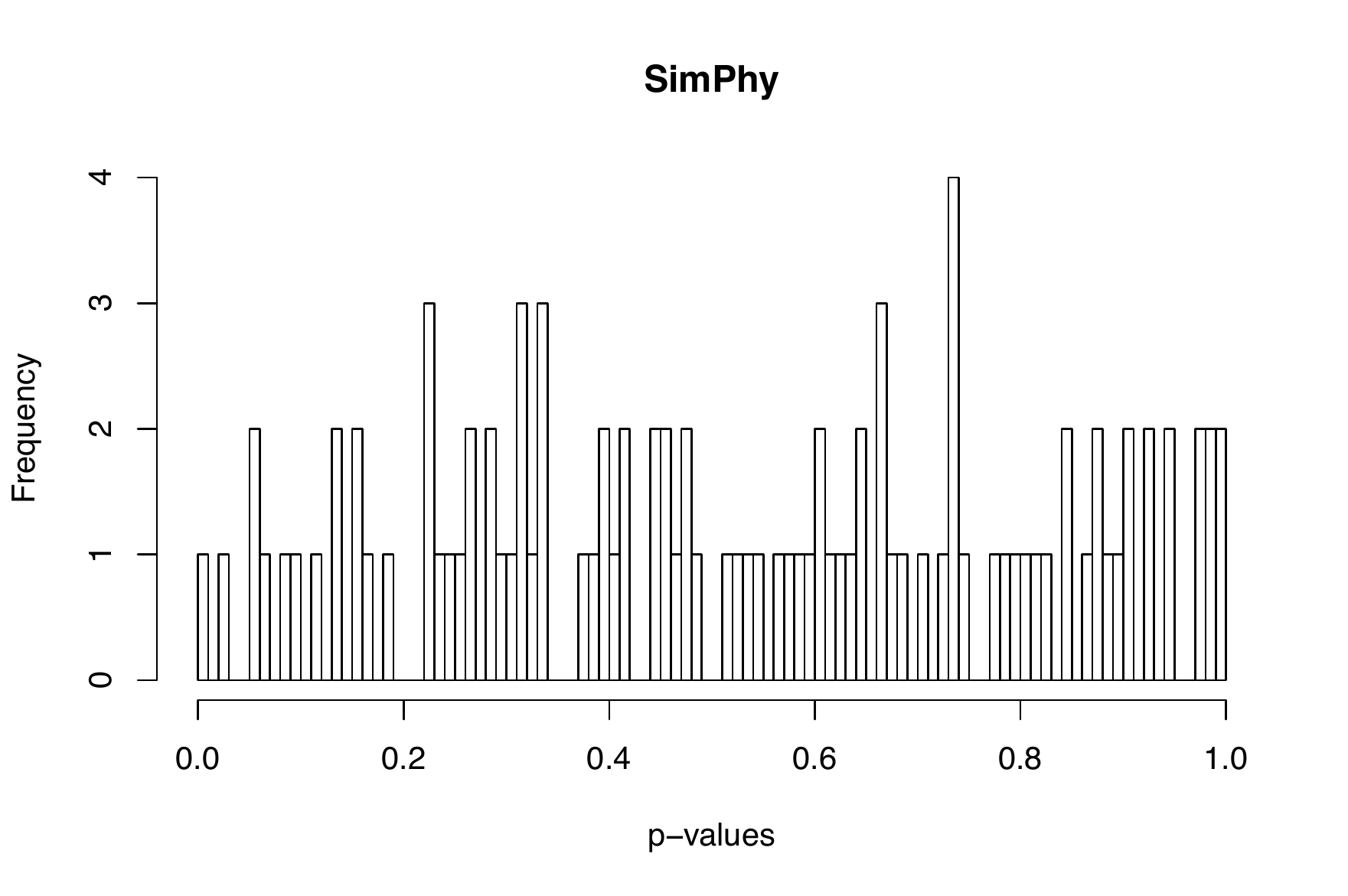}
\includegraphics[width=7.5cm]{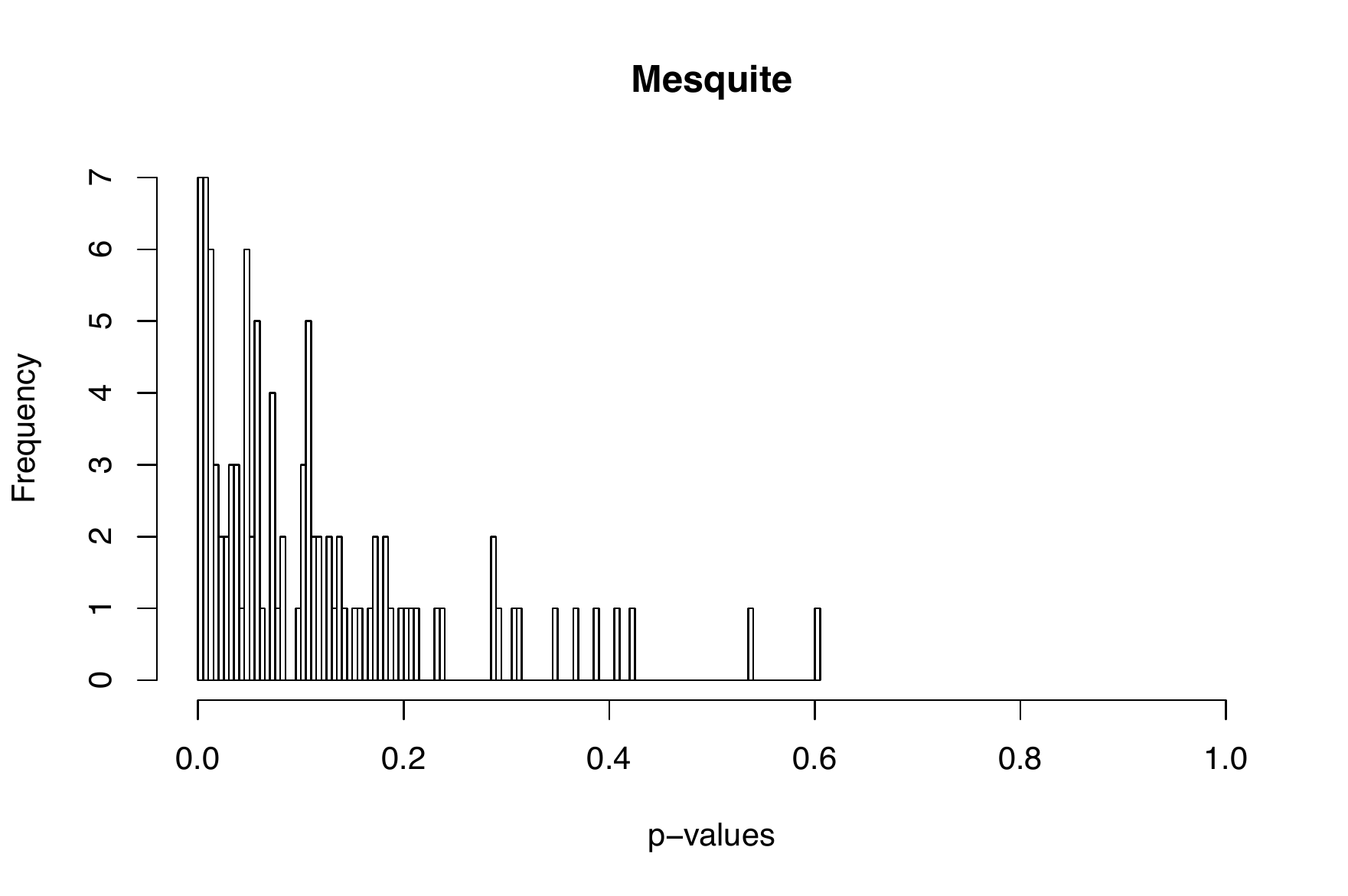}
\end{center}
\caption{Distributions of $p$-values from the Anderson-Darling test
  comparing empirical and theoretical $d(A,B)$ distributions, for
  SimPhy and Mesquite. $p$-values were computed for 100 subsamples of
  size 1000 from the same data underlying Figure
  \ref{fig::pairwise}. The approximate uniformity for SimPhy indicates
  a good fit to theory, while the pronounced skew for Mesquite
  indicates poor fit.} \label{fig::AD}
\end{figure*}

To quantify deviation of the sampled $d(A,B)$ distribution from the
expected distribution, we apply the Anderson-Darling test. Note that
even small numerical errors in a simulation or its analysis may
prevent the extremely close fit to theory that a large sample should
exhibit. As a result, with very large samples such tests can produce
misleadingly small $p$-values, leading to excessive rejection.  We
therefore divided each of our samples into 100 subsamples of size
1000, computing a $p$-value for each. A good fit is shown by a roughly
uniform distribution of $p$-values for the subsamples. Figure
\ref{fig::AD} shows these $p$-value distributions for the samples from
SimPhy and Mesquite, formally confirming the conclusions already
described.

 \begin{table*}
   \caption{Rooted triple topology counts from 100,000 gene trees sampled for the species tree of Figure \ref{fig::simtree} }\label{tab::s3}
 \begin{center} 
 	\begin{tabular}{|c|c|c|c|c|c|}\hline % <-- Alignments: 1st column left, 2nd middle and 3rd right, with vertical lines in between
 		%\multicolumn{5}{|c|}{Topology counts and internal branch estimation  for the triplets of  $S_3$ }\\\hline
 		 &  $((A,B),C)$  & $((A,C),B)$& $((B,C),A)$&$p$-value& Int. branch\\
 		\hline
 		Expected & 59564 &  20217 & 20217 &- & 0.5\\	\hline
 		Mesquite &66000& 17556 &16444  & 0 &0.670 \\	\hline 
 		Hybrid-lambda & 59564 &20195& 20240  &0.975&0.499 \\	\hline
 		Phybase  &  59395 & 20256  & 20349  &  0.492&0.495 \\	\hline
 		SimPhy &  59120  &20397 & 20483 &0.554& 0.497  \\	\hline
 		\multicolumn{6}{|c|}{  }\\\hline
 		 &  $((A,B),D)$  & $((A,D),B)$& $((B,D),A)$&$p$-value& Int. branch\\
 		\hline
 		Expected & 70044& 14977&  14977 &-&$0.83\bar{3}$ \\	\hline
 		Mesquite &74074& 12844& 13082 &4.286e-99& 0.944 \\	\hline 
 		Hybrid-lambda &  70850& 14683& 14466  &0.207&0.827 \\	\hline
 		Phybase  & 71072  & 14450  &14478  &0.940&0.834 \\	\hline
 		SimPhy &   71079  & 14465 & 14456& 0.934& 0.835 \\	\hline      
 		\multicolumn{6}{|c|}{  }\\\hline
 		 &  $((A,C),D)$  & $((A,D),C)$& $((C,D),A)$&$p$-value& Int. branch\\
 		\hline
 		Expected &52231&  23884 &  23884 &-&$0.3\bar{3}$ \\	\hline
 		Mesquite & 50533 &24659 &24808  &6.16e-26&0.298 \\	\hline 
 		Hybrid-lambda &  52230 &23951 &23818  &0.830& 0.3331 \\	\hline
 		Phybase   & 52419  &23606   & 23975  &0.118& 0.337\\	\hline
 		SimPhy &    52141 & 23834 & 24025 &0.579& 0.331 \\	\hline
 		\multicolumn{6}{|c|}{  }\\\hline
 		 &  $((B,C),D)$  & $((B,D),C)$& $((C,D),B)$&$p$-value& Int. branch\\
 		\hline 
 		Expected&52231&  23884 &  23884 &-&$0.3\bar{3}$ \\	\hline
 		Mesquite &  51281 &24389 &24330  &1.32e-08&0.313 \\	\hline 
 		Hybrid-lambda &  52241 &23960 &23798   & 0.758& 0.3335 \\	\hline
 		Phybase   & 52164  & 23825  & 24011  &0.635& 0.331\\	\hline
 		SimPhy &    52220 & 23889 & 23891 &0.997& 0.3330 \\	\hline
 		
 	\end{tabular}
 \end{center}
\end{table*}

\medskip

Topological features of samples were analyzed by tabulating counts of
all rooted triple topologies across the sampled gene trees, and then
performing both a chi-squared test and computing the MLE of the
internal branch length on the species tree triple. Results are shown
in Table \ref{tab::s3}. For all programs except Mesquite, no
$p$-values were extreme enough to suggest poor fit. However,
$p$-values for Mesquite are strongly suggestive of poor fit, being
quite close to 0. Internal branch length estimates were also poorest
for the Mesquite sample. Note that while Hybrid-lambda had poor metric
performance, by this testing procedure it gave a good topological
sample. This behavior is consistent with its pairwise distance
performance when parameters are given in coalescent units; if the
Hybrid-lambda algorithm is based on coalescent units, errors may occur
in conversion into %units of 
numbers of generations.

These results indicate that inadequate attention has been given
previously to ensuring MSC simulators perform correctly. The tests
based on metric and topological summary statistics implemented in the
R package \texttt{MSCsimtester} can uncover errors in simulators, and
in user input to simulators.  We recommend these tests be routinely
used by developers of such simulators and, until software has been
fully vetted, by anyone performing coalescent simulations.

\section{Methods: Derivations of summary distributions} Let
$(S,\{\ell_e\}, \{N_e\})$ be a metric species tree with population
size functions, where each edge $e$ has length $\ell_e$ and population
size $N_e:[0,\ell_e)\to \mathbb R^{>0}$. Here $N_e(t)$ denotes the
population size for a haploid organism $t$ generations above the child
node of $e$. There is also an `above the root' population size
function $N_r:[0,\infty)$. (For diploid taxa, the population sizes
should be doubled). For technical reasons, we assume $1/N_e(t)$ is
integrable on finite intervals.

\subsection{Pairwise distance distribution} 

Let $v$ be the most recent common ancestor of taxa $a$ and $b$ (that
is, the node on $S$ where $A$ and $B$ lineages enter the same
population for the first time), and let $P_a$ be the path in $S$ from
$a$ to $v$, $P_b$ be the path in $S$ from $b$ to $v$, and $P_v$ be the
path in $S$ from $v$ to the root $r$. Then $P_{v}=(e_1,e_2,...,e_k)$,
where $v$ is incident to $e_1$ and $r$ is incident to $e_k$. Finally,
let $g_a=\sum_{e\in P_a} \ell_a$ and $g_b=\sum_{e\in P_b} \ell_b$ be
the number of generations from $a$ and $b$, respectively, to $v$.
Then the distance $d(A,B)$ is a random variable $$Y=g_a+g_b +2X,$$
where $X$ is the random variable giving the time to coalescence of two
lineages at $v$.  Let $c(x)$ be the probability density function for
$X$.

To compute $c(x)$, let $N^*:[0,\infty)\to\mathbb R^{>0}$ be the
piecewise `union' of the $N_e$ for $e\in P$ and $N_r$, which with
$m_0=0$, $m_j=\sum_{i=1}^{j} \ell_{e_i}$ for $1\le j\le k$,
$m_{k+1}=\infty$, and $N_{k+1}$ the population function ancestral to
the root is given by
$$N^*(x)= N_{e_i}(x-m_{i-1})$$
for $x\in [m_{i-1},m_i)$, $1\le i \le k+1$.  Since the coalescent
process for two lineages in the same population of size $N^*(x)$
occurs with instantaneous rate $1/N^*(x)$, the probability density
function is \citep{ALR19}
\begin{align*}
c(x)&=\frac 1{N^*(x)} \exp\left (-\int_0^x    \frac 1{N^*(\tau)} d\tau \right)\\
&=\left (\prod_{j=1}^{i-1} \eta_j\right ) \frac {\exp\left (-\int_0^{x-m_{i-1} }   \frac 1{N_{e_i}(\tau)} d\tau \right)}{N_{e_i}(x-m_{i-1})} ,
\end{align*}
for  $x\in [m_{i-1},m_i)$, where
$$\eta_i=  \exp \left (-\int_0^{\ell_i} \frac 1{N_{e_i}(\tau)} d\tau \right )$$  
is the probability that 2 lineages entering edge $e_i$ fail to coalesce on it.

Since  $X=\frac{Y-g_a-g_b}2$,  setting $g_{ab}=g_a+g_b$ this shows the probability density function for $Y$  is
\[f(y)= \begin{cases} 
0 &\hskip -2.in \text{ for }y \leq g_{ab}, \\
 & \\
\displaystyle \left ( \prod_{j=1}^{i-1} \eta_j \right )\frac{\exp\left(- \int_0^{\frac{y-g_{ab}-2m_{i-1}}2} \frac{1}{N_{e_i}(\tau)}d\tau\right)}  {2N_{e_i}(\frac{y-g_{ab}-2m_{i-1}}2) }& \\
 &\hskip -2.3in  \text{for }\begin{cases}\displaystyle g_{ab}+2m_{i-1}\leq y <  g_{ab}+2m_i,\\ {\displaystyle 1\le i\le k,}\end{cases}\\
 & \\
\displaystyle \left (\prod_{j=1}^k \eta_j \right)\frac{\exp\left(-\int_0^{\frac{y-g_{ab}-2m_{k}}2} \frac{1}{N_{r}(\tau)}d\tau\right)}{2N_{e_r}(\frac{y-g_{ab}-2m_{k}}2) }  & \\
&\hskip -2.in  \text{ for }g_{ab}+2m_k\leq y.\\
\end{cases}
\]
In the special case that a population size function $N_e(t)$ is
constant, this shows that the corresponding piece of $f$ is a shifted,
scaled, and possibly truncated exponential density.

\subsection{Rooted triple frequencies}

Suppose the rooted triple $((a,b),c)$ is displayed on $S$, and let $$P
= (e_1,e_2,\dots e_i)$$ denote the path from the most recent common
ancestor of $a,b$ on $S$ to the most recent common ancestor of
$a,b,c$.  With the notation of the previous subsection, the
probability that the $a$ and $b$ lineages fail to coalesce within $P$
is $\prod_{j=1}^i \eta_j.$ Note that the gene triplets $((a,c),b)$ and
$((b,c),a))$ can only form if the $a,b$ lineages do not coalesce on
$P$. Moreover, if $a,b$ have not coalesced on $P$ then by the
exchangability of lineages in the same populations under the MSC, the
probability that any particular pair of $a,b,c$ coalesce first is 1/3.
Thus $$\mathbb P \left (((a,c),b)\right )= \mathbb P \left ( ((b,c),a)
\right )=\frac 13 \prod_{j=1}^i \eta_j.$$ Since the probabilities of
the three possible topologies sum to 1,
 $$\mathbb P \left ( ((a,b),c) \right )= 1-\frac 23 \prod_{j=1}^i \eta_j.$$
 In the special case of constant population sizes
 $\eta_j=\exp(-\ell_{e_i}/N_{e_i})$. More generally, the length of
 $e_i$ in coalescent units is $\int_0^{\ell_i} \frac 1{N_{e_i}(\tau)}
 d\tau$, and $\prod_{j=1}^i \eta_i =\exp (-x)$ where $x$ is the length
 of $P$ in coalescent units.
 
\section{Acknowledgements}
This research was supported by the National Institutes of Health Grant R01 GM117590, awarded
 under the Joint DMS/NIGMS Initiative to Support Research at the Interface of the Biological and
 Mathematical Sciences. 
 
\bibliography{MSCtest}
\bibliographystyle{natbib}



\section*{Supplementary material}

# Supplementary material: Testing Multispecies Coalescent Simulators using Summary Statistics

Allman, Elizabeth S.
esallman@alaska.edu

Baños, Hector
hbassnos@gmail.com

Rhodes, John A.
j.rhodes@alaska.edu

August 2, 2019

As explained in [1], we used the R package `MSCsimtester` to test four multispecies coalescent (MSC) simulators: Mesquite [3], SimPhy [4], Hybrid-Lambda [5], and Phybase [2]. A sample of 100,000 gene trees was simulated with each of the programs for the following four metric species trees, with one individual sampled per taxon:

$S_1 = ((a{:}1000, b{:}1000){:}1000\#2000, c{:}2000)\#1000$,

$S_2 = (((a{:}1000, b{:}1000){:}1000\#1000, c{:}2000){:}1000\#1000, d{:}3000)\#1000$,

$S_3 = (((a{:}1000, b{:}1000){:}1000\#2000, c{:}2000){:}1000\#3000, d{:}3000)\#1000$, and

$S_4 = (((((a{:}1000, b{:}1000){:}1000\#1000, c{:}2000){:}1000\#3000, d{:}3000){:}1000\#2000, e{:}4000){:}1000\#1000), f{:}5000)\#2000$.

Population sizes are given here for diploid organisms, and the trees are depicted in Figure S1. Caterpillar trees are used as they allow for more edges to be ancestral to the most recent common ancestor of two taxa than any other tree with the same number of taxa. This means that coalescent events for two lineages may occur in more populations. Constant population sizes are used on each species tree edge since the simulators to be tested require that.

While selected results for tree $S_3$ are given in the main paper, more extensive ones are included here.

## 1 Distribution of pairwise distances on gene trees

For each species tree and simulator, and for each pair of taxa, the pairwise distance was extracted from all gene trees in the sample, and a histogram was produced with the theoretical density superimposed, using `MSCsimtester`'s function `pairwiseDist`.

### 1.1 Mesquite

For samples from Mesquite (version 3.5), Figures S2, S3, and S4 show results for species trees $S_1$, $S_2$, and $S_3$ respectively. For $S_4$ we split these histograms between Figures S5 and S6.

1

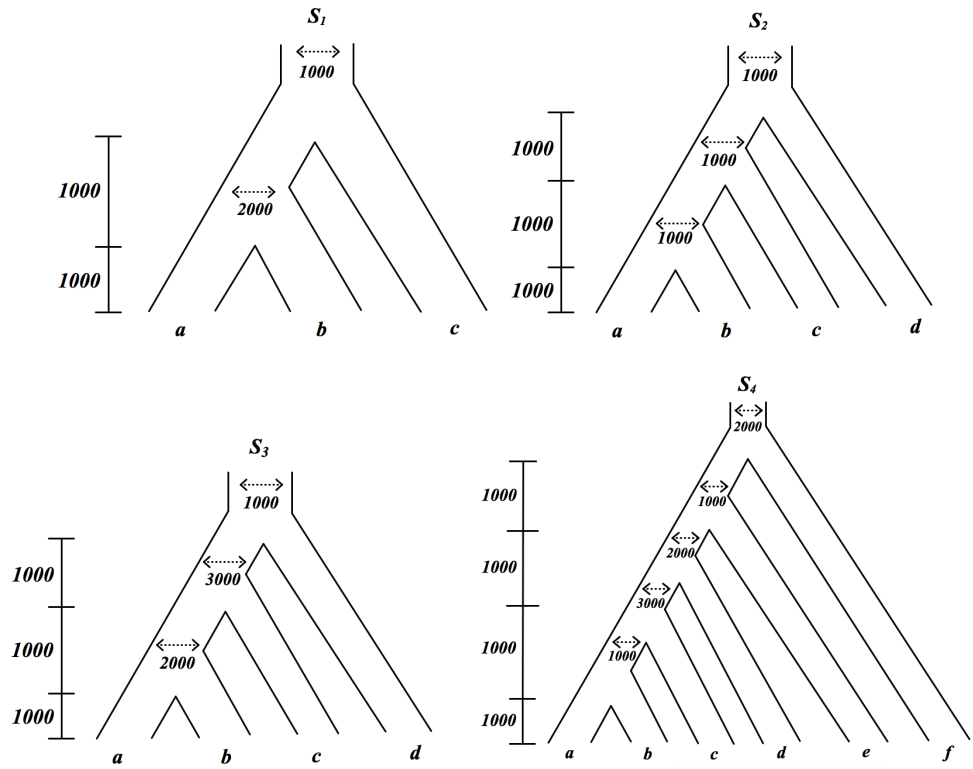

Figure S1: The species trees $S_1$, $S_2$, $S_3$, and $S_4$ used to test multispecies coalescent simulators.

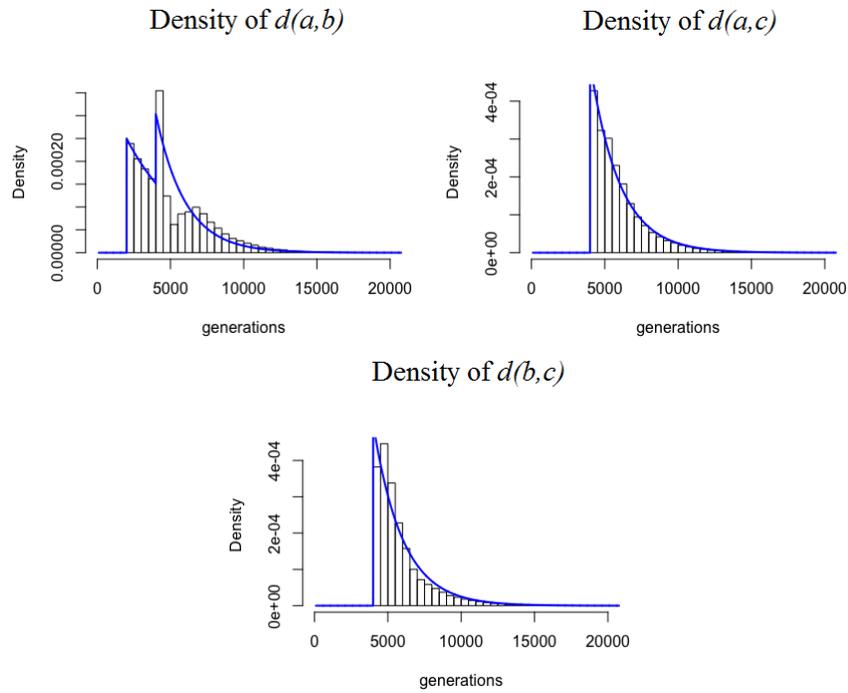

Figure S2: The pairwise gene tree distance probability densities for species on $S_1$, together with the histograms obtained from the distances on the 100,000 gene tree sample simulated by Mesquite.

In Figure S2, the histogram for $d(a,b)$ for the Mesquite sample for $S_1$ exhibits problems due to a population size change at distance 4000. At this point, for the Mesquite sample coalescence occurs too often, followed shortly afterwards by it occurring too seldom to match theory. After this, the incidence of coalescence increases again and leads to an exponential decay similar to that predicted by theory. While the histogram of $d(a,c)$ fits the theoretical distribution better, at around 6000 generations the there is a slight excess in the sampled distances. By the exchangeability property of the coalescent model the histograms of $d(a,c)$ and $d(b,c)$ should match closely, but they do not, as that for $d(b,c)$ shows deficits at roughly 5000 and 7000 generations. Given the large sample size we believe Mesquite does not behave correctly for this species tree.

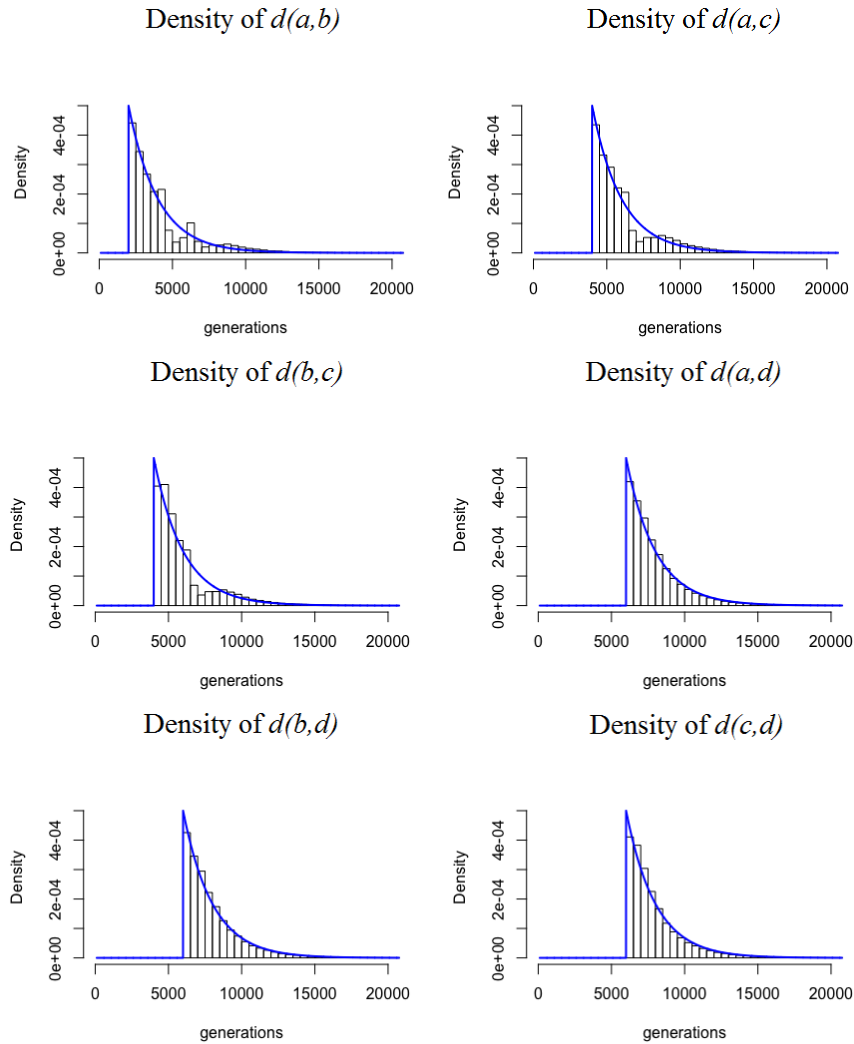

Figure S3: The pairwise gene tree distance probability densities for species on $S_2$, together with the histograms obtained from the distances on the 100,000 gene tree sample simulated by Mesquite.

In Figure S3, which displays the simulated and theoretical density for $S_2$, which has constant population size throughout the tree. We do not see a match in theoretical and simulated empirical distributions in $d(a,b)$, $d(a,c)$, and $d(b,c)$, which should show simple exponential decay. For the remaining pairwise distances the histograms are close to the theoretical distribution. Note the

3

distances showing good fit are the ones with only the population ancestral to the root relevant to the coalescent process.

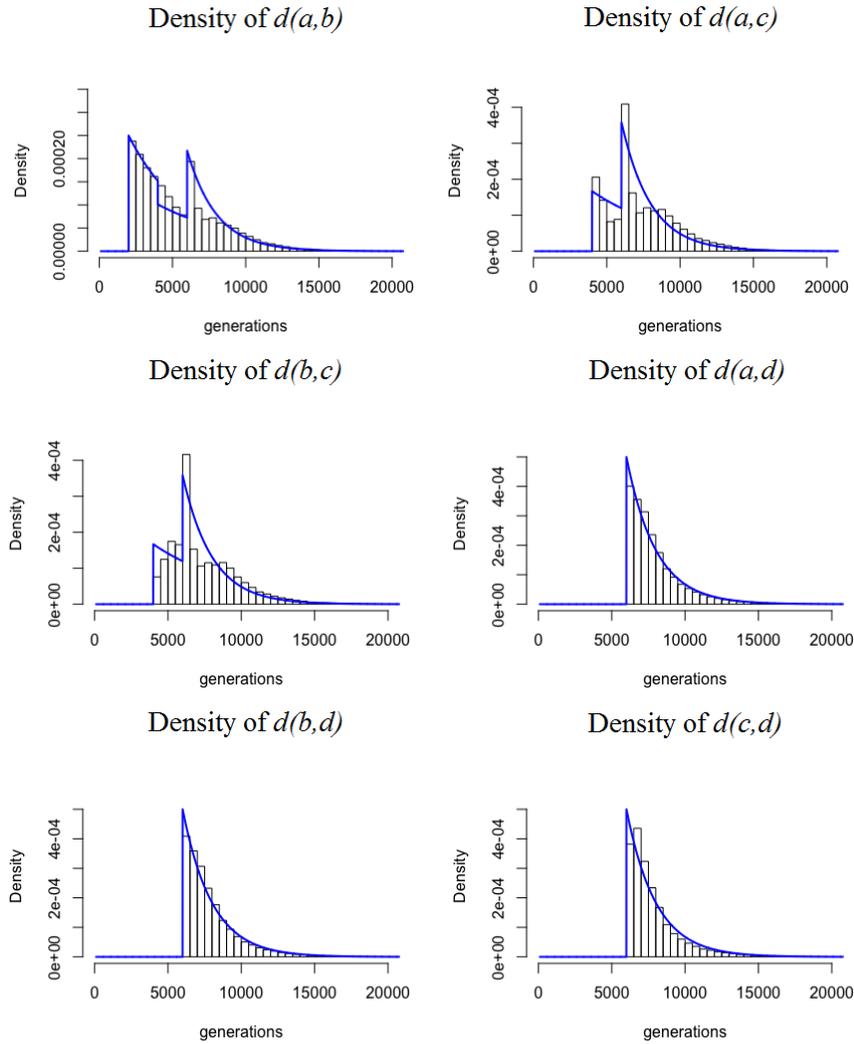

Figure S4: The pairwise gene tree distance probability densities for species on $S_3$, together with the histograms obtained from the distances on the 100,000 gene tree sample simulated by Mesquite.

Turning to Figure S4, the distances histograms for $S_3$ again do not exhibit proper behavior for distances involving several populations, though the remaining histograms behave almost as expected. In Figure S5 and S6 which follow, the theoretical and simulated densities on species tree $S_4$ display similar results.

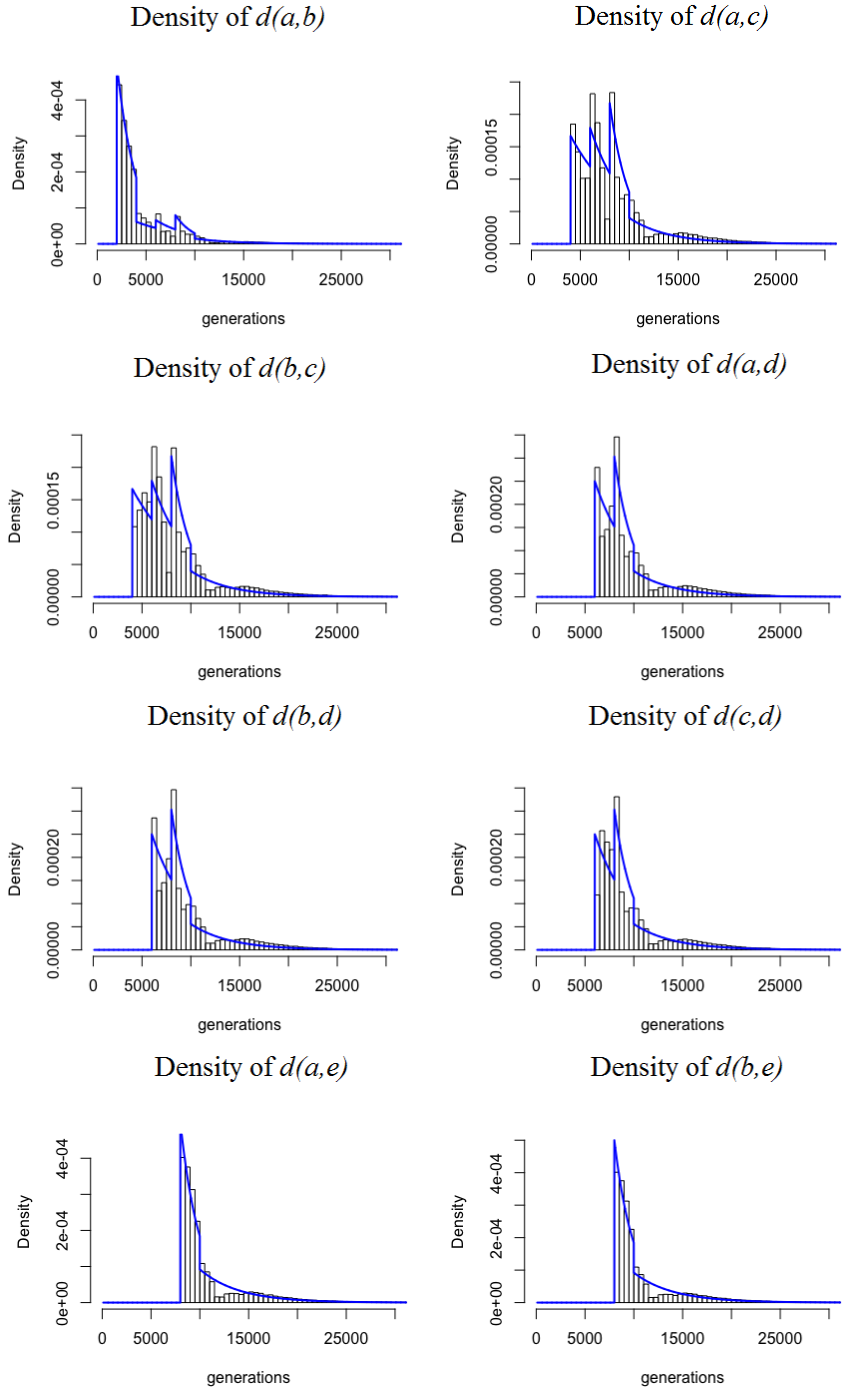

Figure S5: The pairwise gene tree distance probability densities for species on $S_4$, together with the histograms obtained from the distances on the 100,000 gene tree sample simulated by Mesquite.

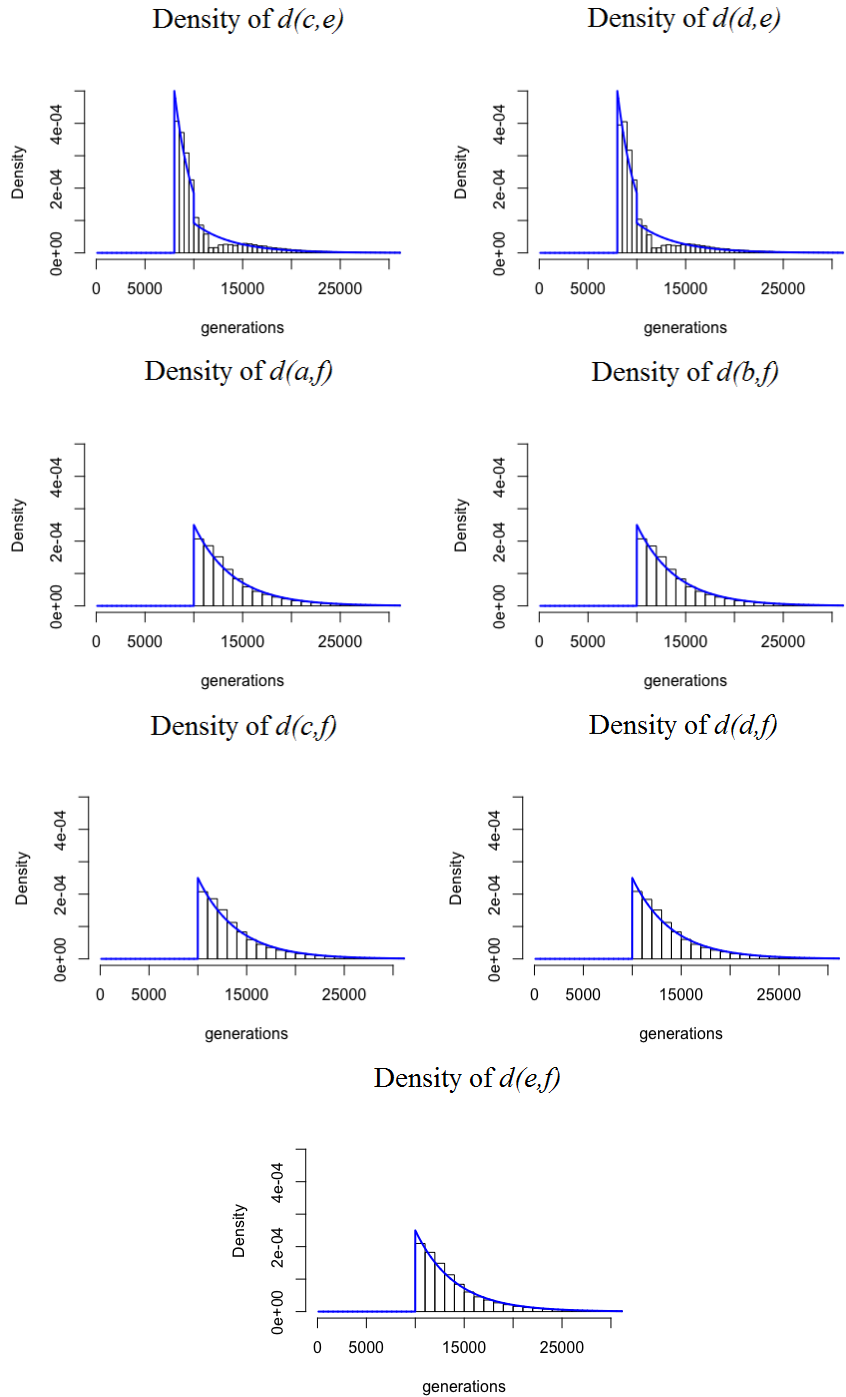

Figure S6: The pairwise gene tree distance probability densities for species on $S_4$, together with the histograms obtained from the distances on the 100,000 gene tree sample simulated by Mesquite.

We conclude that Mesquite simulations do not match the theoretical pairwise distance densities when coalescent events may occur in more than one population on a species tree, regardless of population size or number of species on the tree. The function `ADtest` of the package `MSCsimtester` can be used to perform an Anderson-Darling test on each of the pairwise distances to quantify the poor fit.

## 1.2 Hybrid-Lambda

The tests of Hybrid-Lambda (0.6.1-beta (dev)) are analogous to those conducted with Mesquite. Figures S7, S8, and S9 show histograms for species trees $S_1$, $S_2$, and $S_3$ respectively, with Figures S10 and S11 showing those for $S_4$.

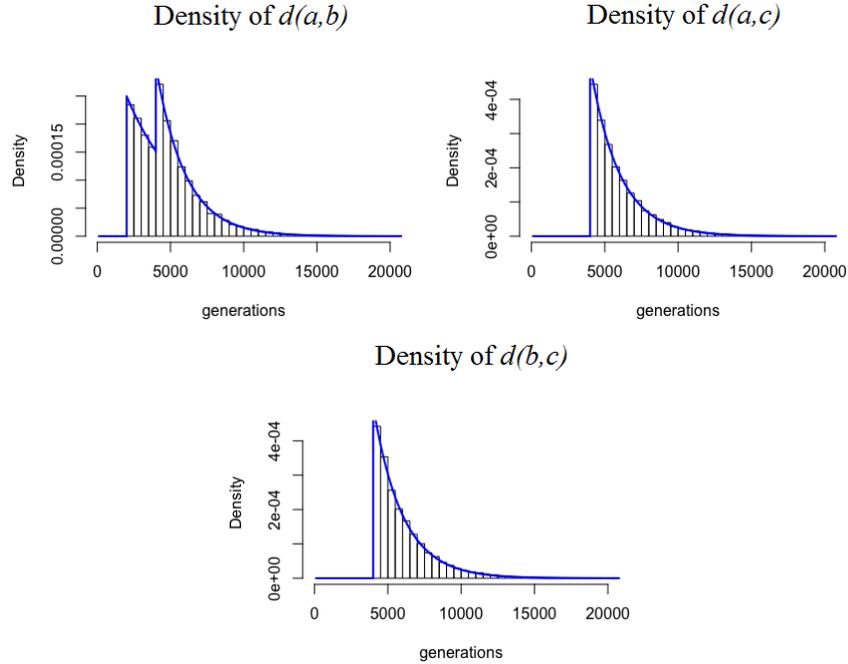

Figure S7: The pairwise gene tree distance probability densities for species on $S_1$, together with the histograms obtained from the distances on the 100,000 gene tree sample simulated by Hybrid-Lambda.

We observe that the Hybrid-Lambda simulations on $S_1$, depicted in Figure S7, show the histograms well approximate the probability densities.

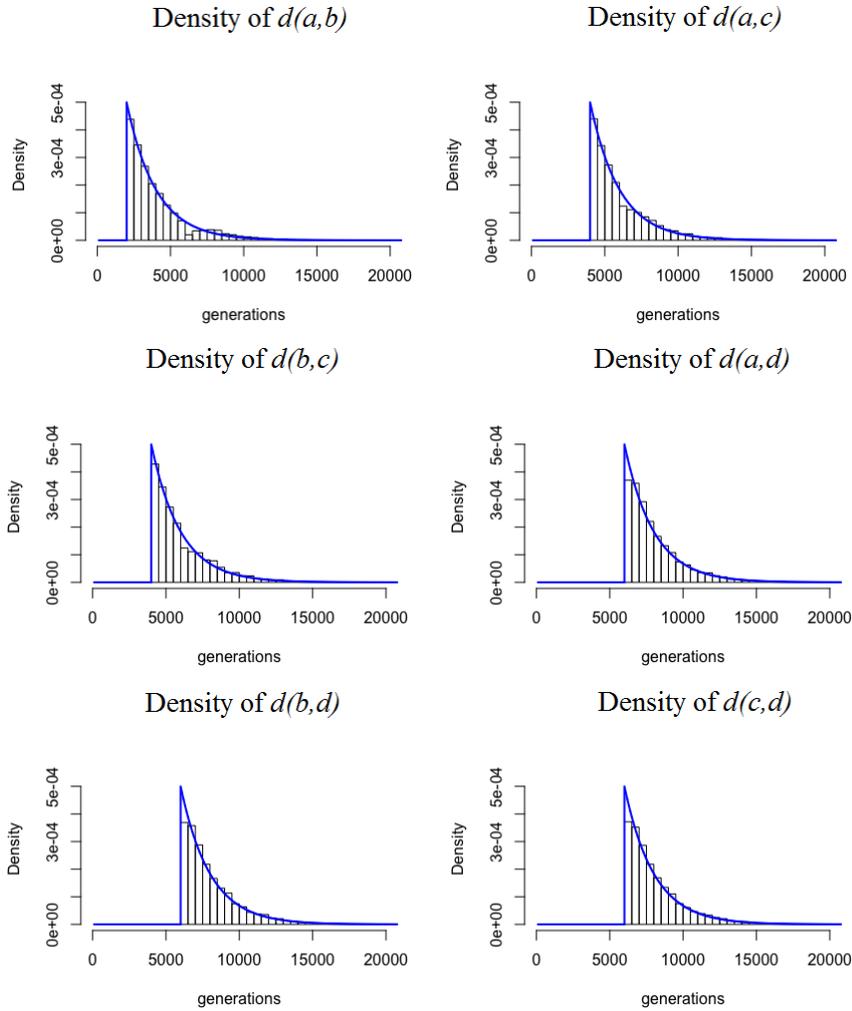

Figure S8: The pairwise gene tree distance probability densities for species on $S_2$, together with the histograms obtained from the distances on the 100,000 gene tree sample simulated by Hybrid-Lambda.

In Figure S8 we observe that the Hybrid-Lambda simulated and theoretical densities of $S_2$ do not match for any distance. The histograms of $d(a,c)$, and $d(b,c)$ are similar to each other (as they should be by exchangeability) but differ with the theoretical density some time after the lineages enter the same population. Likewise, the histograms of $d(a,d)$, $d(b,d)$ and $d(c,d)$ are similar, as exchangeability ensures, but fail to match theory.

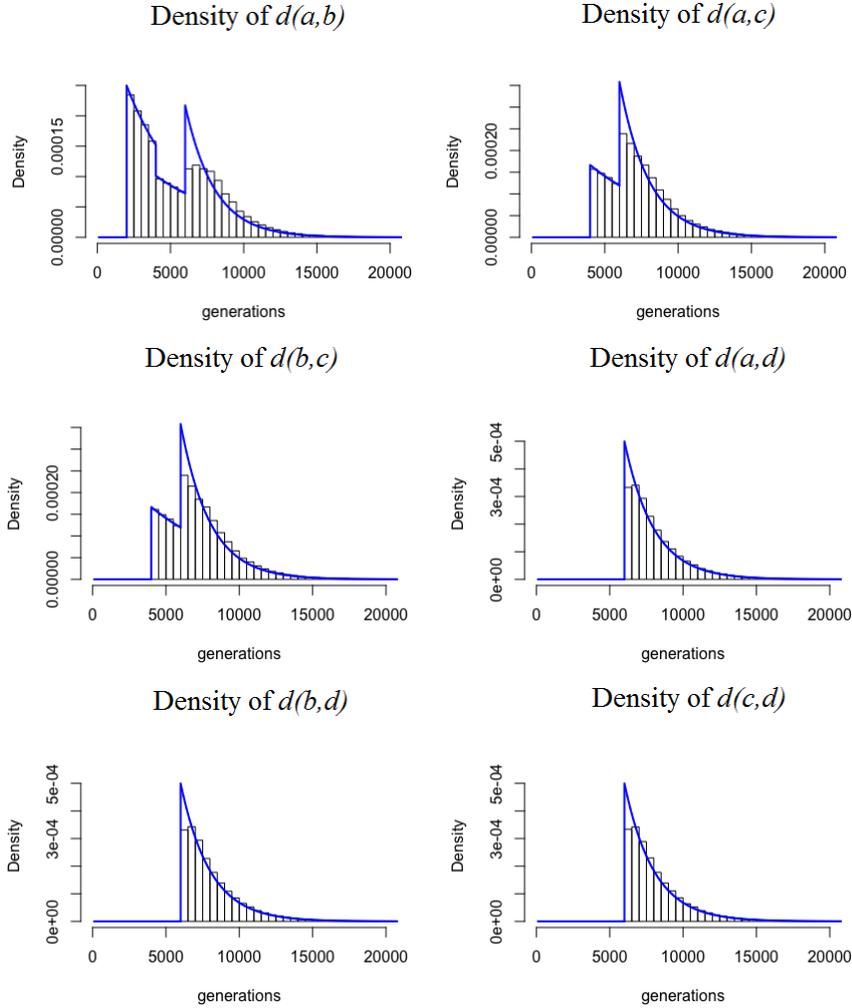

Figure S9: The pairwise gene tree distance probability densities for species on $S_3$, together with the histograms obtained from the distances on the 100,000 gene tree sample simulated by Hybrid-Lambda.

For the Hybrid-Lambda simulation on $S_3$, we observe in Figure S9 a mismatch between all histograms and the theoretical densities. All pairs of lineages show a mismatch when the lineages enter the population at the root. For $S_4$, Figures S10 and S11 show a mismatch in the simulated and theoretical densities for all pairs of taxa. However, for both $S_3$ and $S_4$ simulations the histograms which should agree by exchangeability do so.

9

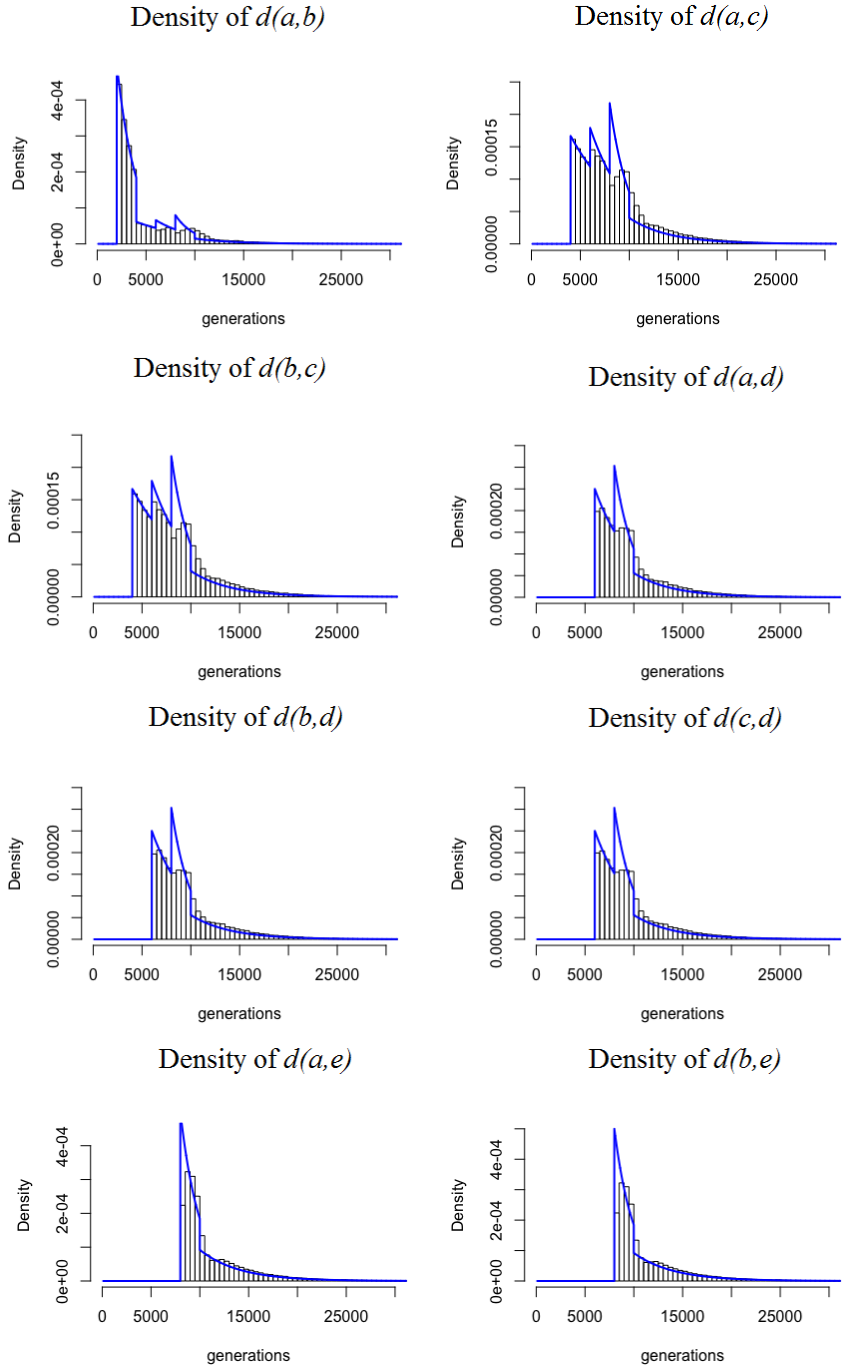

Figure S10: The pairwise gene tree distance probability densities for species on $S_4$, together with the histograms obtained from the distances on the 100,000 gene tree sample simulated by Hybrid-Lambda.

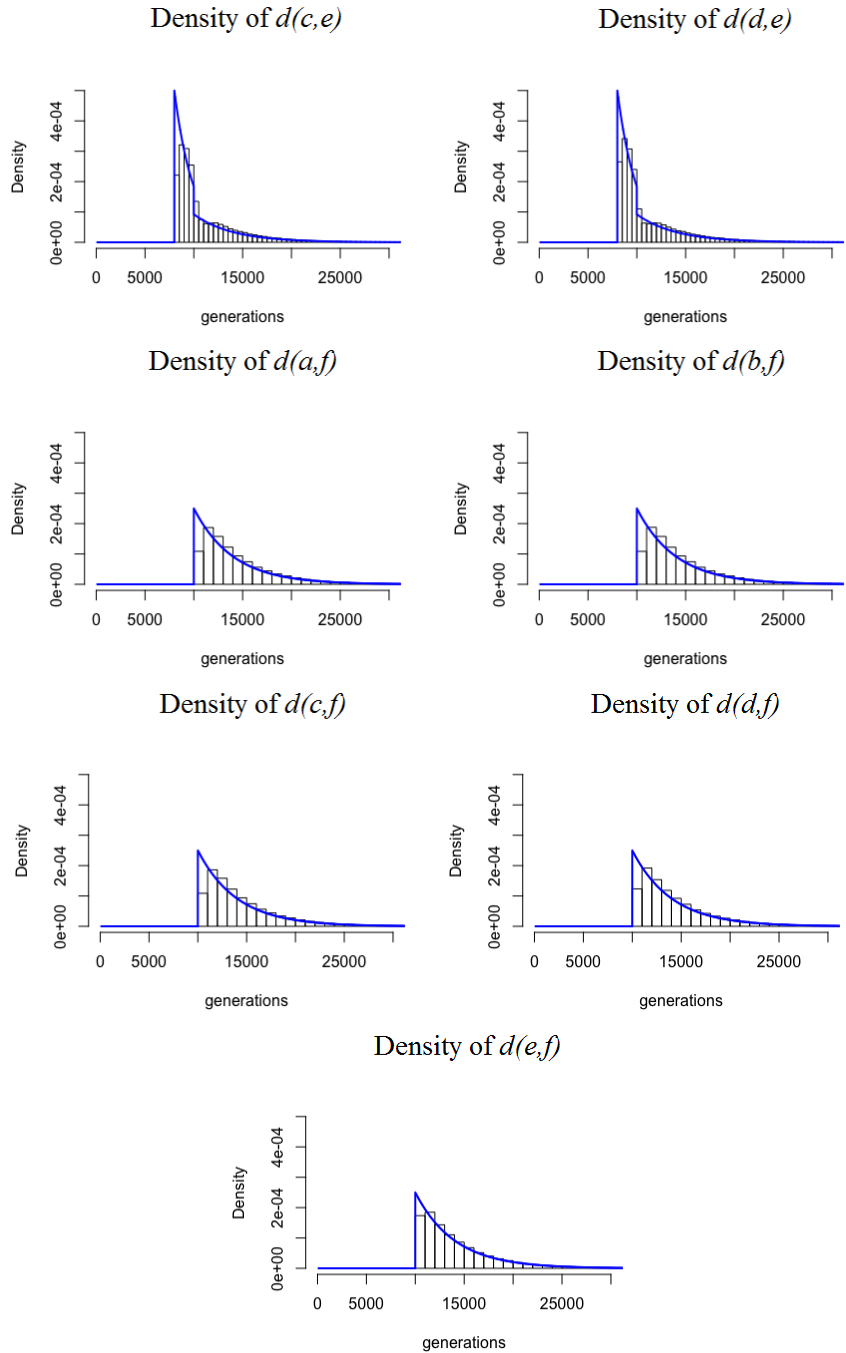

Figure S11: The pairwise gene tree distance probability densities for species on $S_4$, together with the histograms obtained from the distances on the 100,000 gene tree sample simulated by Hybrid-Lambda.

We conclude that Hybrid-Lambda simulations fail to correctly approximate the pairwise distance density for species trees with more than 3 taxa.

## 1.3 SimPhy

The test results for SimPhy are shown in Figures S12, S13, S14, S15 and S16. We observe that for all four trees the histograms closely match theoretical predictions.

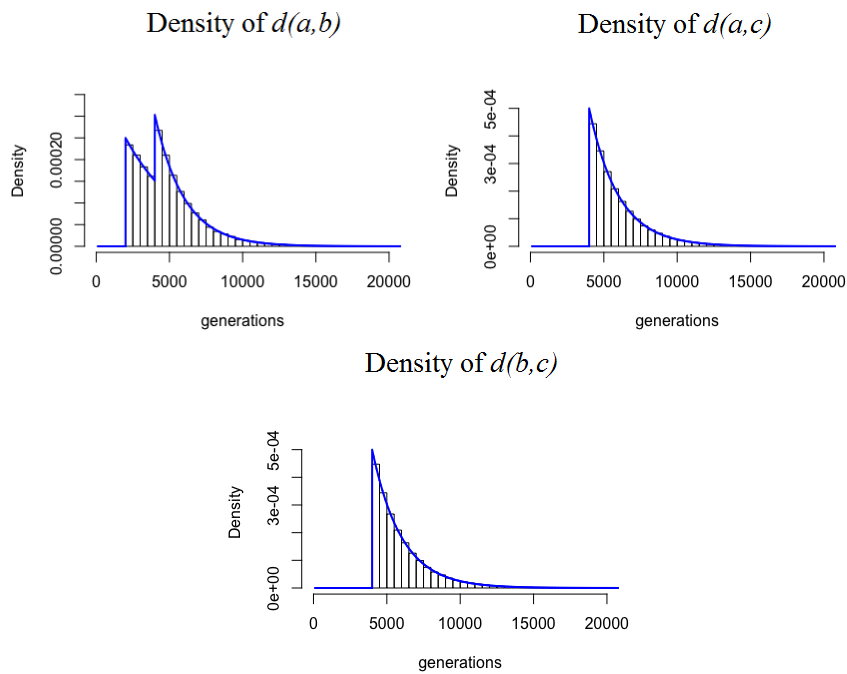

Figure S12: The pairwise gene tree distance probability densities for species on $S_1$, together with the histograms obtained from the distances on the 100,000 gene tree sample simulated by SimPhy.

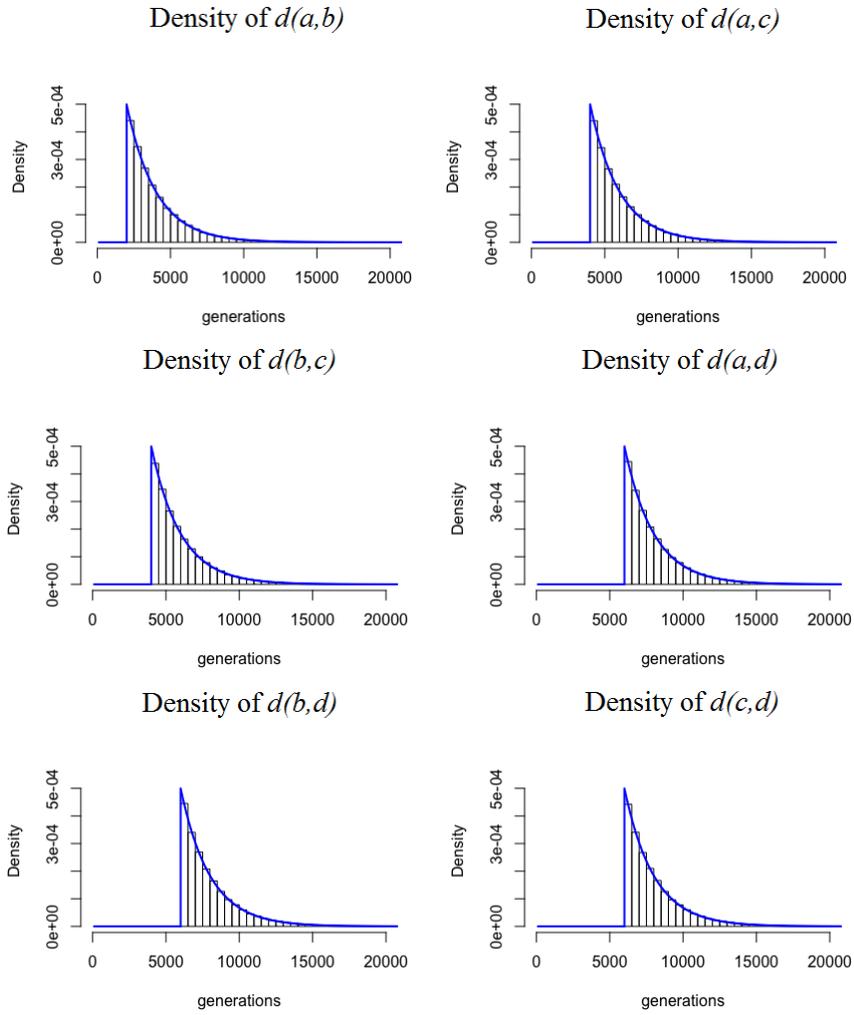

Figure S13: The pairwise gene tree distance probability densities for species on $S_2$, together with the histograms obtained from the distances on the 100,000 gene tree sample simulated by SimPhy.

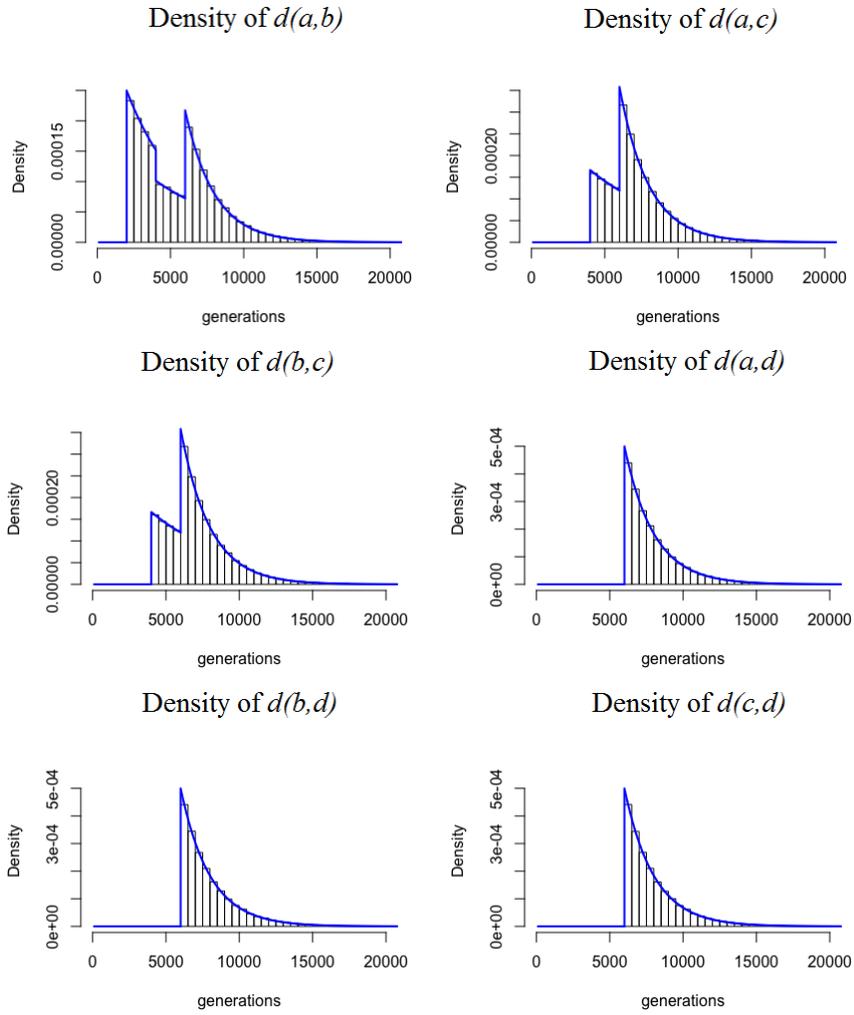

Figure S14: The pairwise gene tree distance probability densities for species on $S_3$, together with the histograms obtained from the distances on the 100,000 gene tree sample simulated by SimPhy.

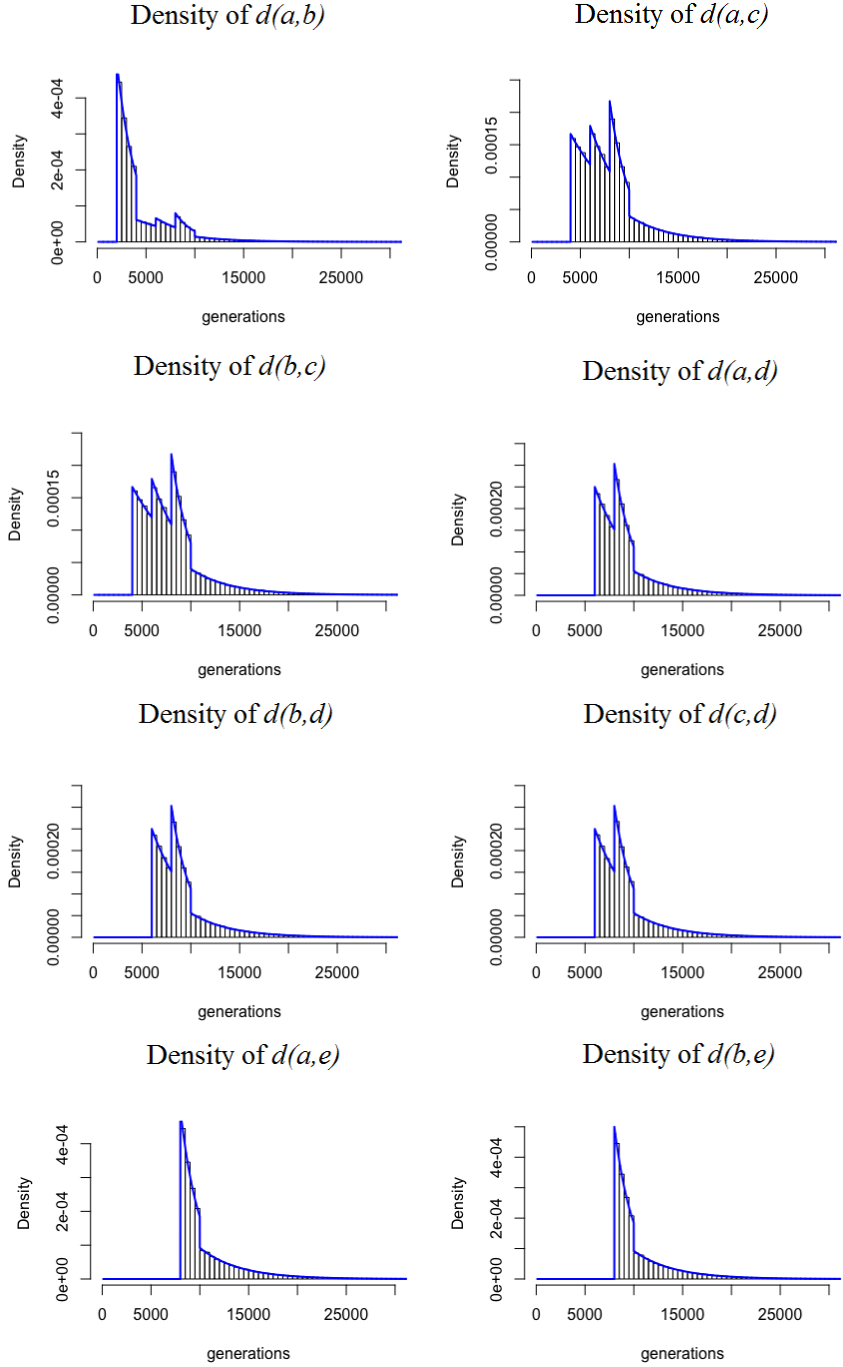

Figure S15: The pairwise gene tree distance probability densities for species on $S_4$, together with the histograms obtained from the distances on the 100,000 gene tree sample simulated by SimPhy.

15

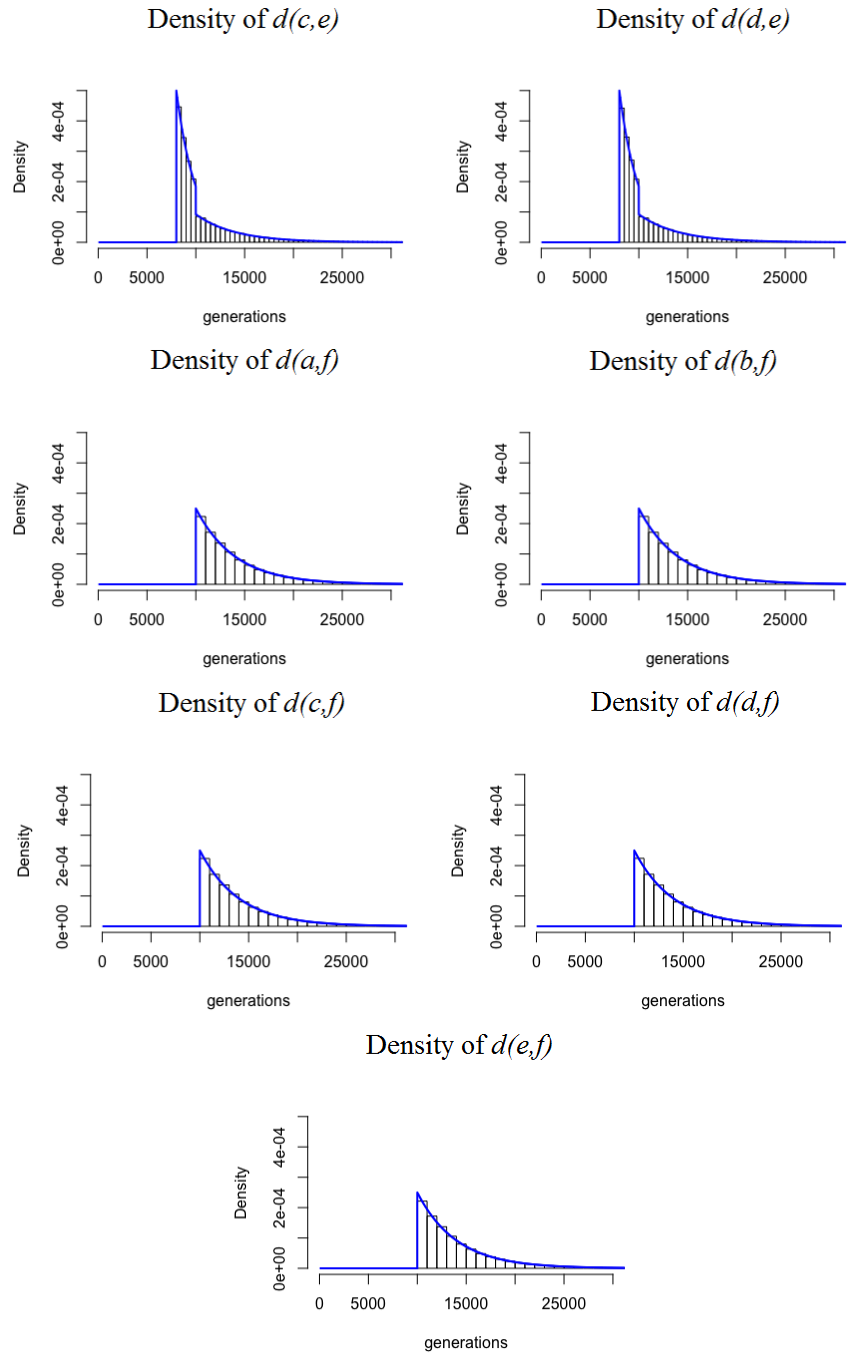

Figure S16: The pairwise gene tree distance probability densities for species on $S_4$, together with the histograms obtained from the distances on the 100,000 gene tree sample simulated by SimPhy.

16

## 1.4 Phybase

Results for simulations with Phybase (version 1.5) are shown in Figures S17, S18, S19, S20 and S21. In all cases there is a good match between the sample and theoretical predictions.

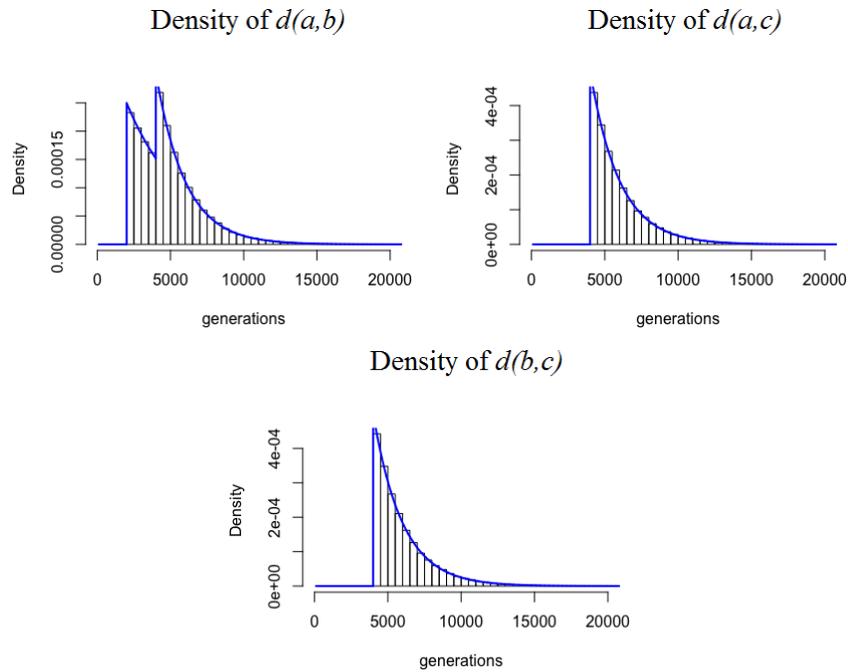

Figure S17: The pairwise gene tree distance probability densities for species on $S_1$, together with the histograms obtained from the distances on the 100,000 gene tree sample simulated by Phybase.

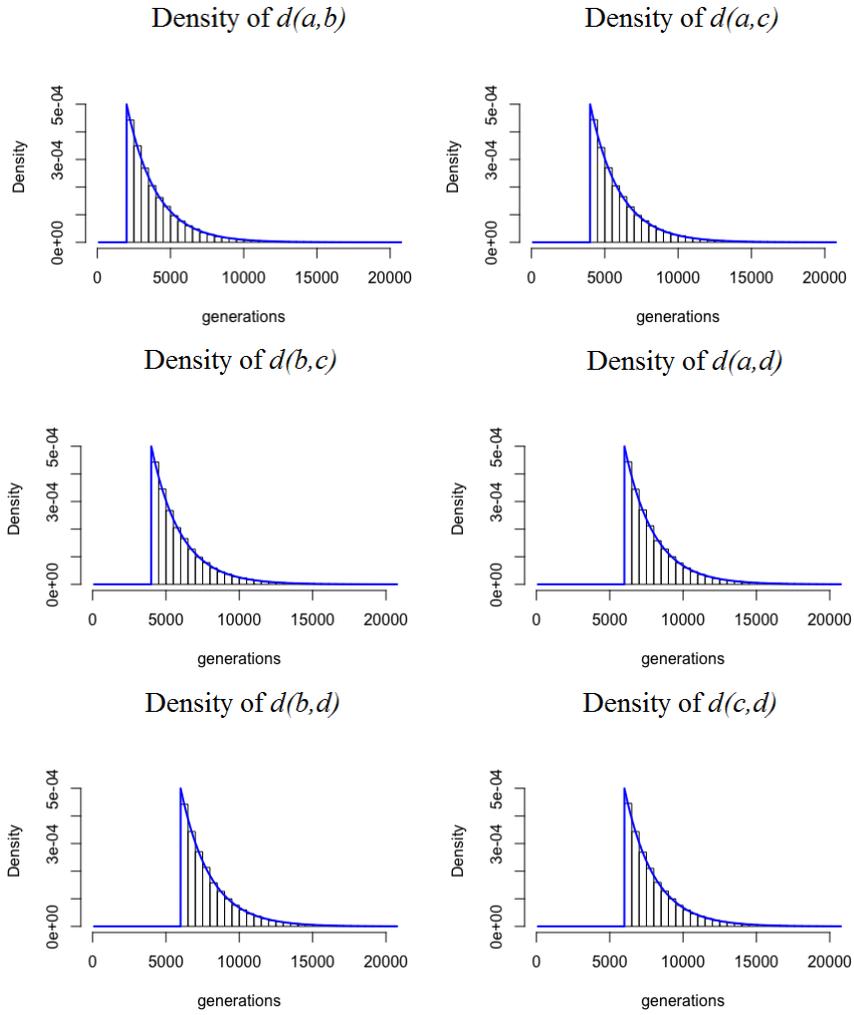

Figure S18: The pairwise gene tree distance probability densities for species on $S_2$, together with the histograms obtained from the distances on the 100,000 gene tree sample simulated by Phybase.

18

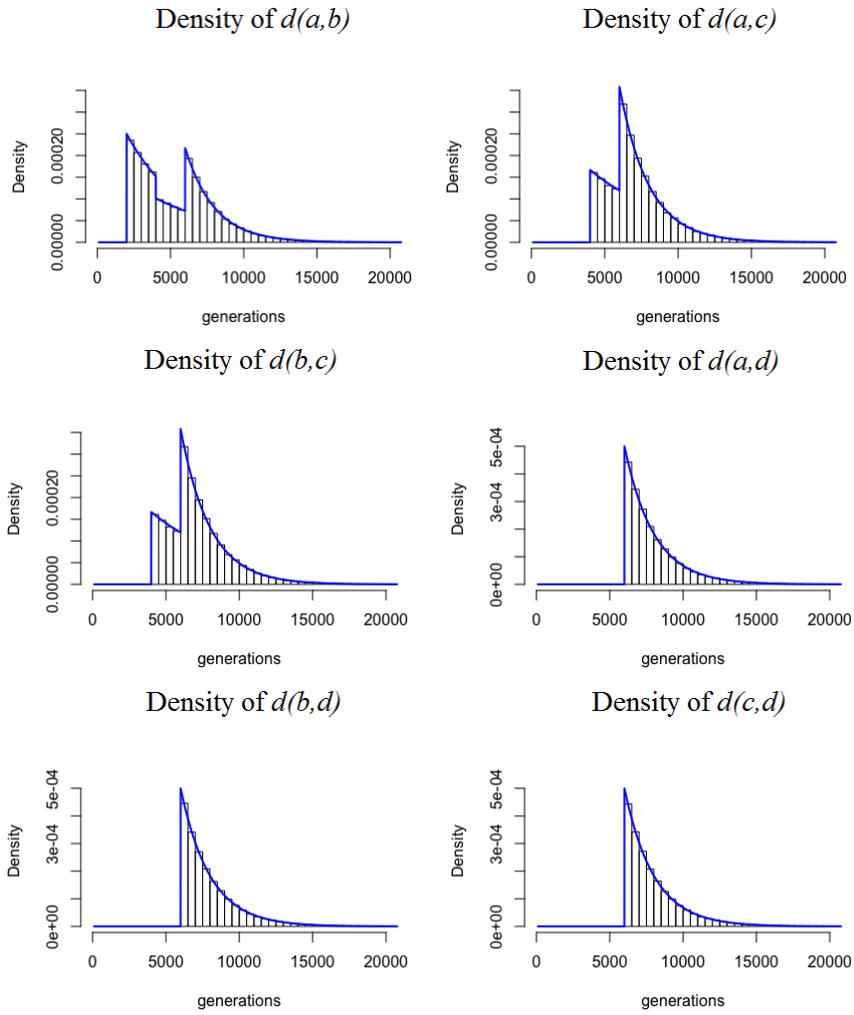

Figure S19: The pairwise gene tree distance probability densities for species on $S_3$, together with the histograms obtained from the distances on the 100,000 gene tree sample simulated by Phybase.

19

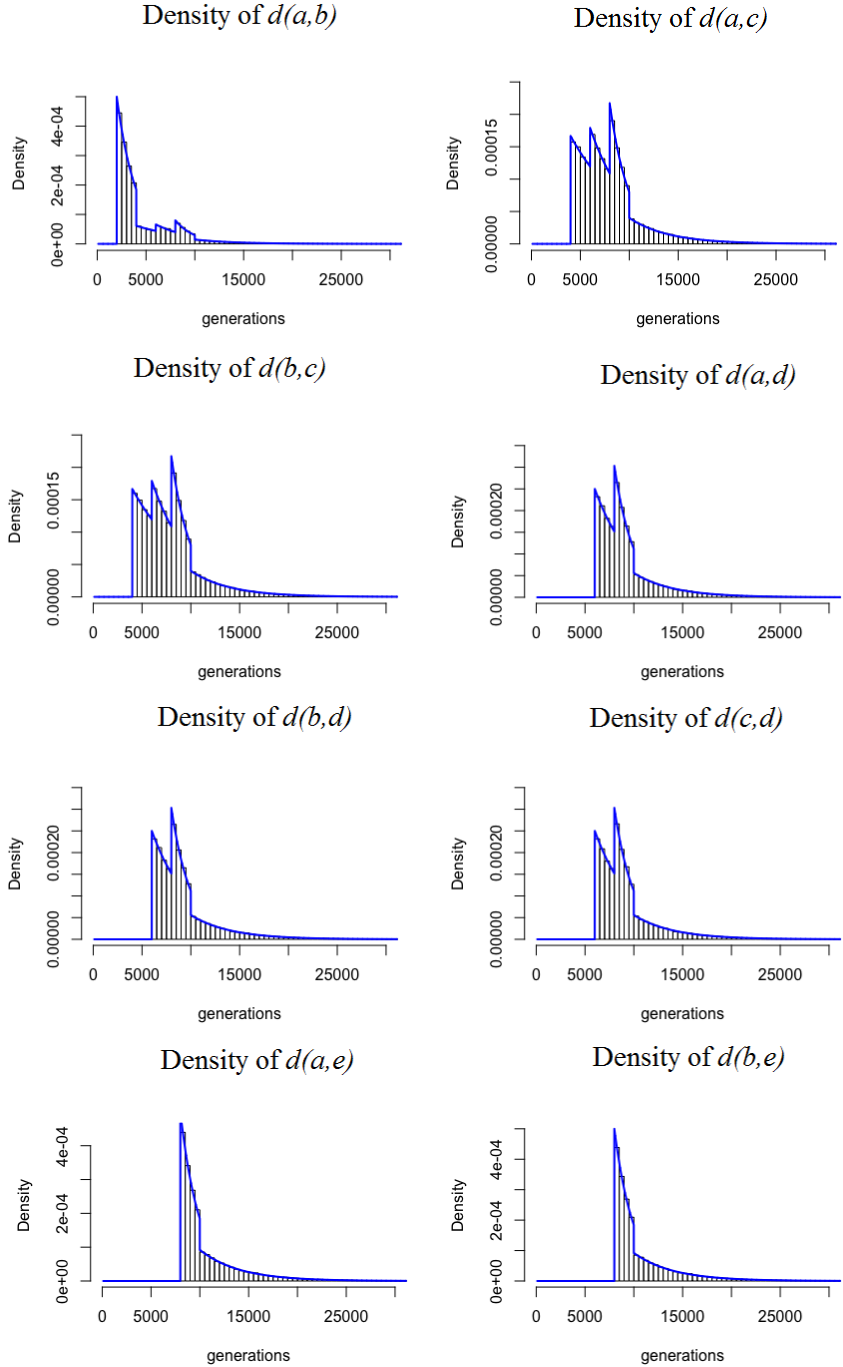

Figure S20: The pairwise gene tree distance probability densities for species on $S_4$, together with the histograms obtained from the distances on the 100,000 gene tree sample simulated by Phybase.

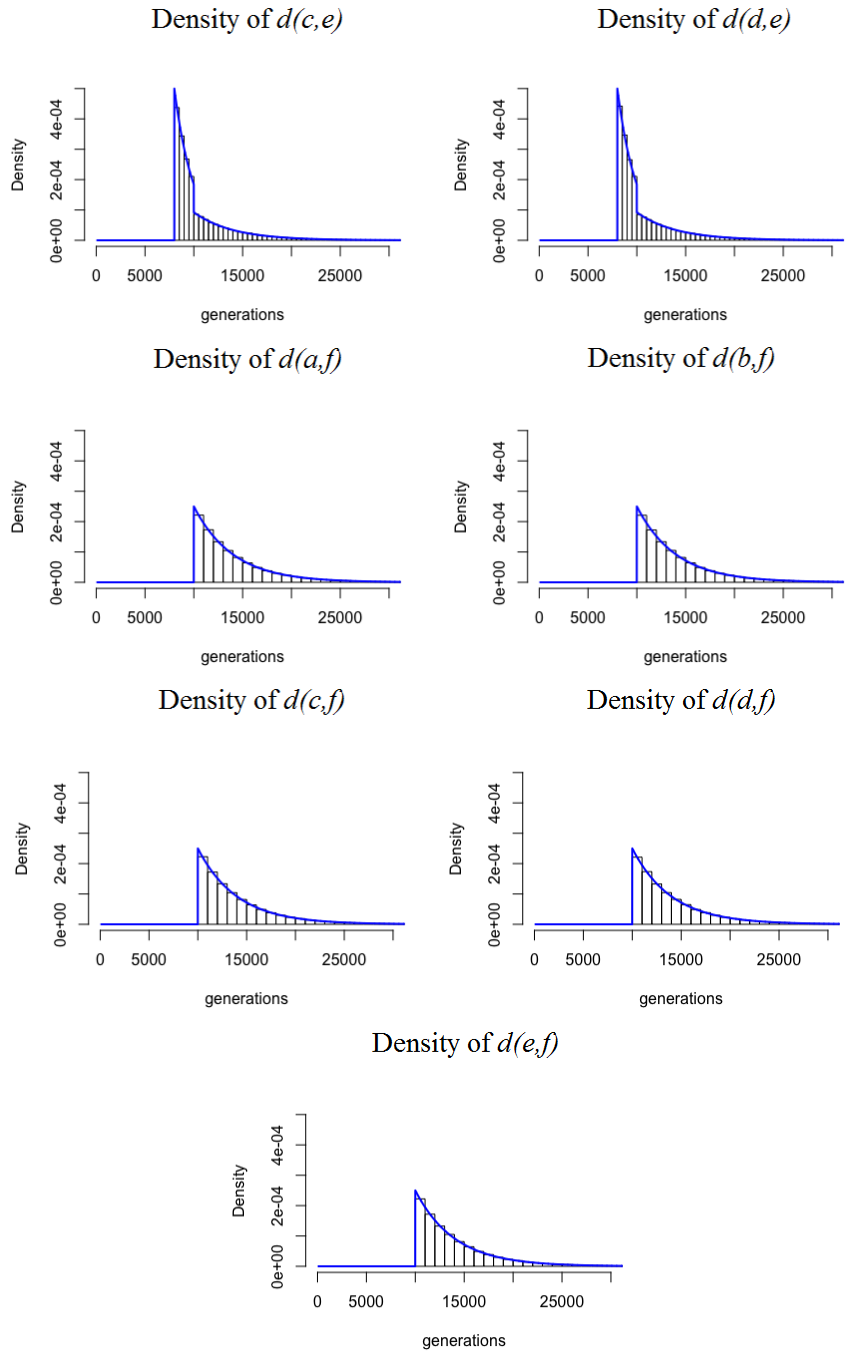

Figure S21: The pairwise gene tree distance probability densities for species on $S_4$, together with the histograms obtained from the distances on the 100,000 gene tree sample simulated by Phybase.

## 2 Distribution of rooted triple topologies on gene trees

The topological features of samples were analyzed by tabulating counts of all rooted triple topologies displayed in the simulation of 100,000 gene trees, using the function `rootedTriple` of the `MSCsimtester` package.

For each of $S_1$, $S_2$, and $S_3$ we examined the set of all possible induced triplets, and for $S_4$ we choose 5 of the $\binom{6}{3} = 20$ induced triplets. For any given triplet in $S_i$ we computed the probability of observing each of the 3 gene tree topologies. We multiplied these probabilities by 100,000 (the sample size) to produce the expected topology counts. We used the theoretical counts and the simulated counts to perform a $\chi^2$-test with two degrees of freedom. For each gene triplet we used this test to obtain a $p$-value. These $p$-values are preliminary results, since this procedure should be repeated on multiple samples to obtain a firmer conclusion. For each rooted triple displayed on the species tree, we computed the internal branch length $\delta$ in coalescent units and compared it to the estimate

$$\hat{\delta} = -\log\left(\frac{3}{2} \cdot \frac{T}{100,000}\right) \quad (1)$$

where $T$ is the count of gene trees in the sample not displaying the induced rooted triple topology.

Tables S1, S2, S3, and S4 show results for the samples on species trees $S_1$, $S_2$, $S_3$, and $S_4$ respectively.

Table S1: Topology counts and branch length estimates for a 100,000 gene tree sample for species tree $S_1$.

| Source | $((A,B),C)$ | $((A,C),B)$ | $((B,C),A)$ | $p$-value | Internal branch |
|---|---|---|---|---|---|
| Theoretical | 59565 | 20218 | 20218 | - | 0.5 |
| Mesquite | 59629 | 20031 | 20340 | 0.281 | 0.501 |
| Hybrid-Lambda | 59511 | 20131 | 20357 | 0.501 | 0.498 |
| Phybase | 59597 | 19999 | 20404 | 0.128 | 0.5008 |
| SimPhy | 59490 | 20289 | 20221 | 0.841 | 0.498 |

In Table S1, we see that all simulators give reasonable estimates of the topology counts and internal branch length for species tree $S_1$. Also, we see no extreme $p$-values for the test of topology counts.

Table S2: Topology counts and branch length estimates for a 100,000 gene tree sample for species tree $S_2$.

| Source | $((A,B),C)$ | $((A,C),B)$ | $((B,C),A)$ | $p$-value | Internal branch |
|---|---|---|---|---|---|
| Theoretical | 75474 | 12262 | 12262 | - | 1 |
| Mesquite | 76009 | 11964 | 12027 | 0.0004 | 1.022 |
| Hybrid-Lambda | 75385 | 12330 | 12284 | 0.770 | 0.996 |
| Phybase | 75224 | 12418 | 12358 | 0.168 | 0.989 |
| SimPhy | 75497 | 12266 | 12237 | 0.970 | 1.0009 |

| Source | $((A,B),D)$ | $((A,D),B)$ | $((B,D),A)$ | $p$-value | Internal branch |
|---|---|---|---|---|---|
| Theoretical | 90977 | 4511 | 4511 | - | 2 |
| Mesquite | 88090 | 5915 | 5995 | 1.609e-221 | 1.723 |
| Hybrid-Lambda | 90834 | 4525 | 4640 | 0.138 | 1.984 |
| Phybase | 90968 | 4521 | 4511 | 0.988 | 1.998 |
| SimPhy | 90915 | 4484 | 4601 | 0.367 | 1.993 |

| Source | $((A,C),D)$ | $((A,D),C)$ | $((C,D),A)$ | $p$-value | Internal branch |
|---|---|---|---|---|---|
| Theoretical | 75474 | 12262 | 12262 | - | 1 |
| Mesquite | 75815 | 12083 | 12102 | 0.044 | 1.013 |
| Hybrid-Lambda | 75574 | 12271 | 12154 | 0.579 | 1.004 |
| Phybase | 75445 | 12353 | 12202 | 0.612 | 0.998 |
| SimPhy | 75305 | 12362 | 12333 | 0.448 | 0.993 |

| Source | $((B,C),D)$ | $((B,D),C)$ | $((C,D),B)$ | $p$-value | Internal branch |
|---|---|---|---|---|---|
| Theoretical | 75474 | 12262 | 12262 | - | 1 |
| Mesquite | 78123 | 10981 | 10896 | 5.09e-83 | 1.114 |
| Hybrid-Lambda | 75579 | 12163 | 12257 | 0.622 | 1.0042 |
| Phybase | 75323 | 12215 | 12462 | 0.153 | 0.993 |
| SimPhy | 75256 | 12444 | 12300 | 0.178 | 0.991 |

In Table S2, for species tree $S_2$, we observe that the Mesquite samples fit theoretical expectation poorly, as shown by the extremely small $p$-values. Specially, we see that some induced rooted triple counts, such as for $((A,D),B)$ and $((B,D),A)$, which should be approximately equal by exchangeability of the model, are not close. While the branch length estimates from the Mesquite samples are in the ballpark of the theoretical values, most are further from them than those of any other simulator. The other simulators appear to produce accurate topological samples.

In Tables S3 and S4 below, the results for $S_3$ and $S_4$ simulations are similar. Mesquite's samples are not in accord with theoretical predictions, while the other simulators' samples are.

23

Table S3: Topology counts and branch length estimates for a 100,000 gene tree sample for species tree $S_3$.

| Source | $((A,B),C)$ | $((A,C),B)$ | $((B,C),A)$ | $p$-value | Internal branch |
|---|---|---|---|---|---|
| Theoretical | 59564 | 20217 | 20217 | - | 0.5 |
| Mesquite | 66000 | 17556 | 16444 | 0 | 0.670 |
| Hybrid-Lambda | 59564 | 20195 | 20240 | 0.975 | 0.499 |
| Phybase | 59395 | 20256 | 20349 | 0.492 | 0.495 |
| SimPhy | 59764 | 20091 | 20145 | 0.504 | 0.422 |

| Source | $((A,B),D)$ | $((A,D),B)$ | $((B,D),A)$ | $p$-value | Internal branch |
|---|---|---|---|---|---|
| Theoretical | 70044 | 14977 | 14977 | - | $0.83\bar{3}$ |
| Mesquite | 74074 | 12844 | 13082 | 4.286e-99 | 0.944 |
| Hybrid-Lambda | 70850 | 14683 | 14466 | 0.207 | 0.827 |
| Phybase | 71072 | 14450 | 14478 | 0.940 | 0.834 |
| SimPhy | 71079 | 14465 | 14456 | 0.934 | 0.835 |

| Source | $((A,C),D)$ | $((A,D),C)$ | $((C,D),A)$ | $p$-value | Internal branch |
|---|---|---|---|---|---|
| Theoretical | 52231 | 23884 | 23884 | - | $0.3\bar{3}$ |
| Mesquite | 50533 | 24659 | 24808 | 6.16e-26 | 0.298 |
| Hybrid-Lambda | 52230 | 23951 | 23818 | 0.830 | 0.3331 |
| Phybase | 52419 | 23606 | 23975 | 0.118 | 0.337 |
| SimPhy | 52323 | 23874 | 23803 | 0.801 | 0.3352 |

| Source | $((B,C),D)$ | $((B,D),C)$ | $((C,D),B)$ | $p$-value | Internal branch |
|---|---|---|---|---|---|
| Theoretical | 52231 | 23884 | 23884 | - | $0.3\bar{3}$ |
| Mesquite | 51281 | 24389 | 24330 | 1.32e-08 | 0.313 |
| Hybrid-Lambda | 52241 | 23960 | 23798 | 0.758 | 0.3335 |
| Phybase | 52164 | 23825 | 24011 | 0.635 | 0.331 |
| SimPhy | 52326 | 23843 | 23831 | 0.843 | 0.3353 |

Table S4: Topology counts and branch length estimates for a 100,000 gene tree sample for species tree $S_4$.

| Source | $((A,B),F)$ | $((A,F),B)$ | $((B,F),A)$ | $p$-value | Internal branch |
|---|---|---|---|---|---|
| Theoretical | 96078 | 1960 | 1960 | - | $2.83\bar{3}$ |
| Mesquite | 95535 | 2275 | 2190 | 3.01e-18 | 2.703 |
| Hybrid-Lambda | 96035 | 1959 | 2005 | 0.590 | 2.822 |
| Phybase | 96055 | 1997 | 1948 | 0.677 | 2.827 |
| SimPhy | 96034 | 1987 | 1979 | 0.749 | 2.821 |

| Source | $((A,C),F)$ | $((A,F),C)$ | $((C,F),A)$ | $p$-value | Internal branch |
|---|---|---|---|---|---|
| Theoretical | 89341 | 5329 | 5329 | - | $1.83\bar{3}$ |
| Mesquite | 86271 | 6866 | 6863 | 8.77e-216 | 1.58 |
| Hybrid-Lambda | 89459 | 5336 | 5204 | 0.212 | 1.844 |
| Phybase | 89392 | 5304 | 5304 | 0.876 | 1.838 |
| SimPhy | 89403 | 5304 | 5293 | 0.817 | 1.839 |

| Source | $((B,C),F)$ | $((B,F),C)$ | $((C,F),B)$ | $p$-value | Internal branch |
|---|---|---|---|---|---|
| Theoretical | 89341 | 5329 | 5329 | - | $1.83\bar{3}$ |
| Mesquite | 86583 | 6666 | 6751 | 1.88e-174 | 1.6 |
| Hybrid-Lambda | 89298 | 5330 | 5371 | 0.838 | 1.829 |
| Phybase | 89256 | 5314 | 5430 | 0.361 | 1.825 |
| SimPhy | 89362 | 5320 | 5318 | 0.978 | 1.835 |

| Source | $((A,D),F)$ | $((A,F),D)$ | $((D,F),A)$ | $p$-value | Internal branch |
|---|---|---|---|---|---|
| Theoretical | 85124 | 7437 | 7437 | - | 1.5 |
| Mesquite | 80531 | 9764 | 9705 | 0 | 1.23 |
| Hybrid-Lambda | 85152 | 7540 | 7307 | 0.156 | 1.501 |
| Phybase | 85181 | 7382 | 7434 | 0.800 | 1.503 |
| SimPhy | 84902 | 7500 | 7598 | 0.100 | 1.485 |

| Source | $((C,D),E)$ | $((C,E),D)$ | $((D,E),C)$ | $p$-value | Internal branch |
|---|---|---|---|---|---|
| Theoretical | 59564 | 20217 | 20217 | - | 0.5 |
| Mesquite | 61541 | 19212 | 19247 | 6.25e-36 | 0.55 |
| Hybrid-Lambda | 59631 | 20282 | 20086 | 0.567 | 0.501 |
| Phybase | 59890 | 20075 | 20035 | 0.109 | 0.508 |
| SimPhy | 59701 | 20256 | 20043 | 0.389 | 0.503 |



\end{document}